\documentclass[sn-mathphys-num]{sn-jnl}

\usepackage{graphicx, subfigure}%
\usepackage{multirow}%
\usepackage{amsmath,amssymb,amsfonts}%
\usepackage{amsthm}%
\usepackage{mathrsfs}%
\usepackage[title]{appendix}%
\usepackage{xcolor}%
\usepackage{textcomp}%
\usepackage{manyfoot}%
\usepackage{booktabs}%
\usepackage{algorithm}%
\usepackage{algorithmicx}%
\usepackage{algpseudocode}%
\usepackage{listings}%

\theoremstyle{thmstyleone}%
\theoremstyle{thmstyletwo}%

\theoremstyle{thmstylethree}%

\begin{document}

\title[Mitigating Interference of Microservices with a Scoring Mechanism in Large-scale Clusters]{Mitigating Interference of Microservices with a Scoring Mechanism in Large-scale Clusters}


\author[1]{\fnm{Dingyu} \sur{Yang}}\email{dingyu.ydy@alibaba-inc.com}

\author[2]{\fnm{Kangpeng} \sur{Zheng}}\email{zheng.k.p@sjtu.edu.cn}

\author*[2]{\fnm{Shiyou} \sur{Qian}}\email{qshiyou@sjtu.edu.cn}

\author[2]{\fnm{Jian} \sur{Cao}}\email{cao-jian@sjtu.edu.cn}

\author[2]{\fnm{Guangtao} \sur{Xue}}\email{gt\_xue@sjtu.edu.cn}

\affil[1]{\orgdiv{TRE}, \orgname{Alibaba Group}, \orgaddress{\street{Chuanhe Road 55}, \city{Shanghai}, \postcode{201203},  \country{China}}}

\affil[2]{\orgdiv{Computer Science and Engineering}, \orgname{ Shanghai Jiao Tong University}, \orgaddress{\street{Dongchuan Road 800}, \city{Shanghai}, \postcode{200240},  \country{China}}}


\abstract{Co-locating latency-critical services (LCSs) and best-effort jobs (BEJs) constitute the principal approach for enhancing resource utilization in production. Nevertheless, the co-location practice hurts the performance of LCSs due to resource competition, even when employing isolation technology. 
Through an extensive analysis of voluminous real trace data derived from two production clusters, we observe that BEJs typically exhibit periodic execution patterns and serve as the primary sources of interference to LCSs. Furthermore, despite occupying the same level of resource consumption, the diverse compositions of BEJs can result in varying degrees of interference on LCSs. Subsequently, we propose PISM, a proactive Performance Interference Scoring and Mitigating framework for LCSs through the optimization of BEJ scheduling. Firstly, PISM adopts a data-driven approach to establish a characterization and classification methodology for BEJs. Secondly, PISM models the relationship between the composition of BEJs on servers and the response time (RT) of LCSs. Thirdly, PISM establishes an interference scoring mechanism in terms of RT, which serves as the foundation for BEJ scheduling. We assess the effectiveness of PISM on a small-scale cluster and through extensive data-driven simulations. The experiment results demonstrate that PISM can reduce cluster interference by up to 41.5\%, and improve the throughput of long-tail LCSs by 76.4\%.}




\maketitle

\section{Introduction}
\label{sec:introduction}

Latency-critical services (LCSs) and best-effort jobs (BEJs) are co-located on the same server to optimize resource utilization in various production clusters  \cite{borg20} \cite{colocatingworkload} \cite{Dirigent} \cite{Prophet} \cite{PARTIES} \cite{CLITE} \cite{PerfIso}. The resource utilization of LCSs is typically associated with the workload, commonly measured in terms of queries per second (QPS). Due to the periodic nature of LCSs' workload, their resource usage exhibits significant fluctuations \cite{PerfIso}. To enhance the overall resource utilization, low-priority BEJs are scheduled alongside LCSs. BEJs are characterized by their computationally intensive nature and relaxed requirements for response time (RT), often involving multiple instances.

However, co-location introduces performance interference issues arising from resource competition among applications \cite{heracles} \cite{CoPart} \cite{mindthegap}. Interference undermines resource efficiency and poses a significant challenge to all applications, particularly LCSs with stringent RT requirements.
Analyzing and mitigating interference in production clusters is a complex task, which manifests in three key aspects.
Firstly, co-located applications exhibit diverse resource requirements and intensity, leading to distinct performance behaviors. 
Secondly, interference is reciprocal, making it intricate to identify the sources of interference among tens to hundreds of co-located applications on the same server.
Thirdly, production clusters experience high levels of dynamism as applications and resource usage constantly change.

When addressing interference concerns in production clusters hosting co-located applications, the predominant emphasis lies on the interference encountered by LCSs owing to the imperative need to ensure their QoS. While LCSs may potentially face interference from other LCSs and BEJs sharing the same server, it is noteworthy that LCSs are typically assigned to specific CPU cores in production \cite{borg2020}, with BEJs being the main resource consumers. Consequently, researchers mainly investigate the impact of BEJs on LCSs in terms of interference.

Existing literature on interference analysis and mitigation usually focuses on deploying specific combinations of applications and predicting their runtime performance in limited-scale scenarios. These studies typically pre-characterize applications under the influence of controlled-intensity interference \cite{heracles} \cite{Bubble-up} \cite{Paragon}, thereby limiting their applicability in production environments. 
Moreover, certain investigations employ feedback-based methods to mitigate interference among co-located applications \cite{PARTIES} \cite{CLITE}, which involves adjusting LCS resources based on QoS requirements with a few seconds of delay. However, these methods face challenges when applied in highly dynamic production clusters.



We conduct an in-depth analysis on a real datasets derived from our production cluster comprising 10,000 servers equipped with Intel(R) Xeon(R) Platinum 8269CY CPUs and boasting 512GB of memory. In addition, we analyze the Google dataset which is obtained from a cluster encompassing 12,000 servers. We acquire valuable observations that facilitate a deeper comprehension of application characteristics, the sources of interference, and the performance of various LCSs on interference. Firstly, BEJs dominate the resource usage of
servers and are the source of LCS performance interference. Secondly, BEJs have high repeatability in large-scale clusters. Thirdly, despite occupying the same level of resource consumption, the diverse compositions of BEJs can result in varying degrees of interference on LCSs.

Considering the highly dynamic nature of clusters, we delve into the characterization of BEJs and extract their collective representation, referred to as server BEJ compositions, to assess their interference on LCSs.
On one hand, although individual instances of BEJ consume fewer resources and have short durations, the presence of numerous co-located BE instances in each server leads to significant resource consumption. For example, our production cluster witnesses the execution of approximately 60 new BE instances per minute on each server. Consequently, measuring the interference generated by each BE instance in such a highly dynamic environment becomes impracticable.
On the other hand, certain implicit patterns observed in production clusters, such as the notable repeatability of BEJs and the relative stability of LCSs, provide opportunities to address the interference problem.

Utilizing a data-driven approach, we propose PISM, an innovative framework for scoring and mitigating performance interference among co-located applications in production clusters. Firstly, PISM adopts a data-driven approach to establish a characterization and classification methodology for BEJs. Secondly, PISM models the relationship between the composition of BEJs on servers and the response time (RT) of LCSs. Thirdly, PISM establishes an interference scoring mechanism in terms of RT, which serves as the foundation for BEJ scheduling.

We extensively evaluate the effectiveness and performance of PISM using a small-scale cluster comprising 14 servers, along with large-scale data-driven simulations. The experiment results demonstrate that PISM can reduce cluster interference by up to 41.5\%, and improve the throughput of long-tail LCSs by 76.4\%.

Our main contributions are summarized as follows:

\begin{itemize}

    
     \item \textbf{An interference scoring and mitigating framework.} Building on observations, we devise PISM, a lightweight, non-invasive, and easily integrable framework that effectively scores and mitigates interference. PISM aims to predict interference before scheduling BEJs.
    
    \item \textbf{A characterization and classification method for BEJs.} To simplify interference analysis, we propose a method to characterize BEJs based on resource usage and classify them into distinct categories. This approach enables a shift in perspective from analyzing individual BEJs to examining BEJ compositions.
    
    
    \item \textbf{Two interference scoring models.} We develop a scoring model to assess the degree of interference experienced by individual LCSs, effectively reflecting their performance. Additionally, we establish an interference scoring model for servers by assigning appropriate weights to different LCSs executing on the same server.

\end{itemize}

The remainder of this paper is organized as follows: Section \ref{sec-back} describes the background. Section \ref{sec-related-work} discusses the related work. Section \ref{sec_ob} provides observations. Section \ref{sec_system} details the design of PISM. Section \ref{sec_evaluation} presents the experiment results. We conclude the work in Section \ref{sec-conclusion}.

\section{Background}
\label{sec-back}

In order to enhance the self-contained and readability of this work, we introduce some background of microservice colocation.

\subsection{LC Services} 
LCSs are some applications with latency-critical requirements, that handle user requests with strict real-time responses, exemplified by tasks like web searches, instant messaging, and e-commerce\cite{Dirigent}. These systems are continuously to promptly address incoming requests. 
To handle the dynamic workload, LCSs typically deploy numerous instances across various servers. Load balancing is employed to distribute requests uniformly among these LCS instances. The number of LCS instances is not static but is adjusted in response to fluctuations in incoming requests \cite{autopilot}\cite{hyscale}. 

In the scheduling process of microservices, multiple instances of one LCS are typically deployed on different servers to ensure fault tolerance, preventing the disruption of specific instances caused by hardware failures. Intuitively, each instance of the same microservice is expected to exhibit similar performance, given that they have identical workloads. However, due to their execution in the servers with intricate colocation environments and varying degrees of interference, their performance may fluctuate considerably and frequently, which has been investigated in both Google Clusters \cite{googleworkload} and Alibaba Clusters \cite{colocatingworkload, taskdependencies}. This variability poses a challenge in accurately measuring and quantifying these dynamic changes in production.

\begin{figure}[tbp]
\centering
\includegraphics[width=0.85\linewidth]{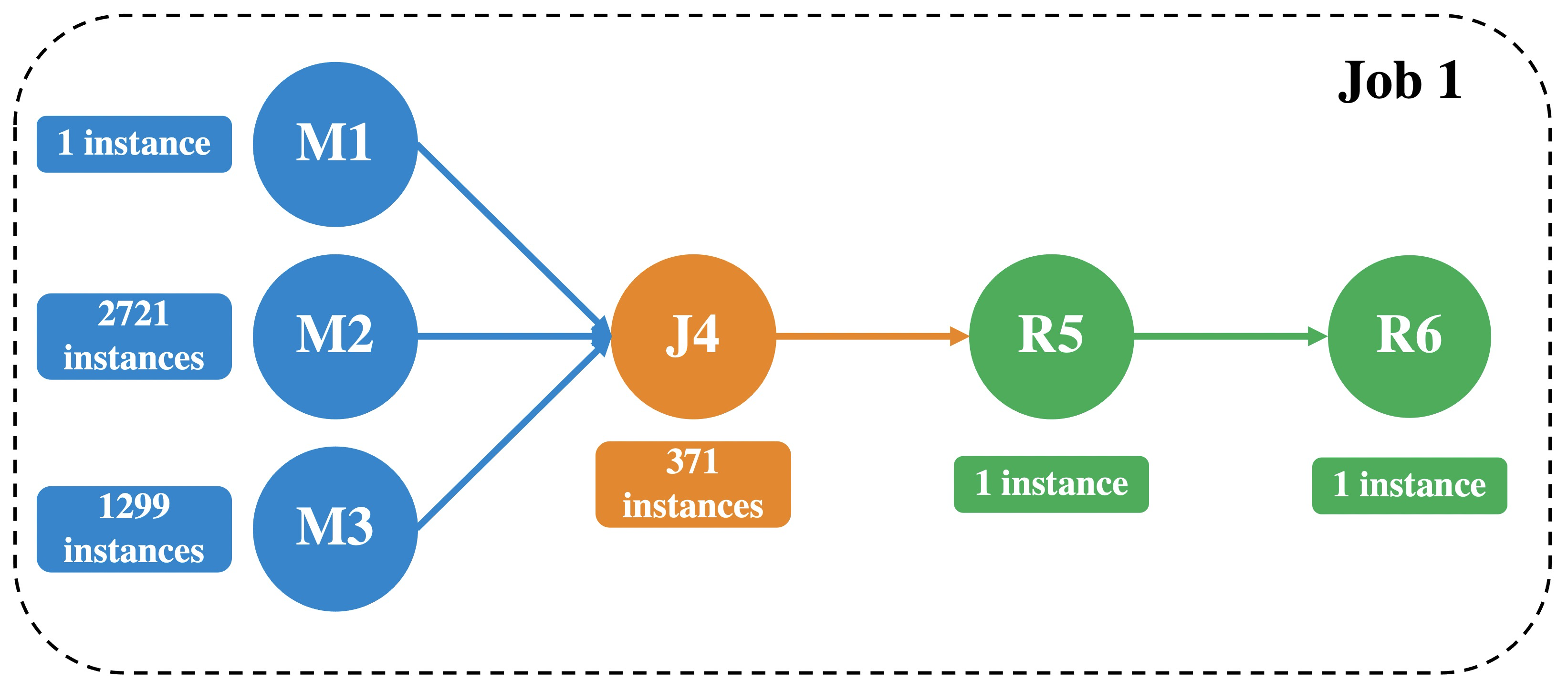} 
\caption{The task composition of an example BEJ.} 
\label{DAG} 
\end{figure}

\subsection{Best-Effort Jobs} 
Best-Effort Jobs (BEJs) refer to the batch jobs designed to periodically process large-scale, repetitive data jobs without any explicit Quality of Service (QoS) guarantees or sensitive latency requirements. 
Most BEJs involve data processing and are generally organized into three hierarchical levels: job, task, and instance. Specifically, a BEJ may comprise multiple tasks executed sequentially following their designated stages within the job. There may be dependencies between tasks in a BEJ \cite{taskdependencies} \cite{Unearthing}. This relationship is often represented by a directed acyclic graph (DAG). The DAG model is conventionally utilized in large-scale data processing systems like MapReduce or Spark. For example, Figure \ref{DAG} illustrates the task composition of an example BEJ with six tasks, comprising three map tasks, one join task and two reduce tasks. 
In a single BE task, the quantity of its instances is determined by the volume of the data to be processed. 

Instances are the main body of BEJs and the smallest units for scheduling. Typically, BE instances are not bound to particular CPU cores; instead, they can be executed on any available cores, efficiently improving CPU utilization.
However, since BE instances and LCS instances are colocated in the same server, they inevitably share some computing resources (such as CPU Caches). This incurs some potential resource contention and impacts the Quality of Service (QoS) for LCSs. 

In this work, we analyze the performance interference from the LCS perspective to ensure the QoS of LCSs. We understand that there may be times when certain components, such as BEJs, may need to be sacrificed or rescheduled to maintain production efficiency.

\section{Related Work}
\label{sec-related-work}

\subsection{Application Characterization.} 
Due to the complexity and variability of applications in production, it is not easy to directly model their performance. To reduce complexity, application or workload characterization is usually the first step by analyzing private or open-source datasets and running benchmarks to extract the patterns of applications. The workloads studied can be divided into some categories, such as co-located applications \cite{borg20} \cite{colocatingworkload} \cite{wholimit}, LCSs or microservices\cite{heracles}  \cite{SoftSKU} \cite{MicroserviceCharacterizing} \cite{DeathStarBench} \cite{ workloadcharacter}, BEJs \cite{taskdependencies} \cite{googleworkload}, and machine learning jobs \cite{PAI} \cite{DistributedDL} \cite{DNNScheduler} \cite{MLaaS}.

\subsection{Performance Interference Modeling} 
Numerous studies have been conducted to investigate the correlation between the operating environment and complex application co-locations, aiming to predict the performance in the face of dynamic workloads, resource contention, and various interference. In order to evaluate the interference, some work \cite{Prophet} \cite{heracles} adopt response time to denote the QoS metric and design an interference model to predict the performance degradation precisely. 
Several performance interference models introduce architecture-level metrics such as CPI and MIPS to measure performance \cite{Measuring} \cite{DeepDive} \cite{CPI2} \cite{Bubble-flux}.
However, obtaining application-level or architecture-level performance metrics requires some effort or designing sophisticated tools to collect the performance data, which is sometimes not feasible. 

Currently, there is no unified standard to identify interference sources and quantify resource competition pressure. The work in \cite{ibench} summarizes 15 interference sources and proposes iBench, a tool that can generate multi-dimensional performance interference for artificial interference injection. \cite{Bubble-up} and \cite{Bubble-flux} use a memory-tunable program to simulate different levels of memory interference, while \cite{Seer} introduces the length of a request queue within the service as pressure. \cite{Paragon} and \cite{Mage} also consider the impact of heterogeneous clusters on interference models.

\subsection{Co-located Cluster Scheduling and Resource Management} 
When co-locating LCSs and BEJs to improve resource utilization and efficiency, some careful scheduling strategies need to be considered for BEJs to minimize interference among applications, especially on LCSs. Traditionally, predictive-aware scheduling models can be utilized to estimate the application performance after co-location, enabling the identification and filtering of servers with the highest interference levels \cite{Prophet} \cite{Bubble-up} \cite{Paragon} \cite{Mage} \cite{Quasar} \cite{Quincy}. 
Another approach \cite{ResourceCentral} \cite{Hawk} \cite{kang2022hwoa} involves predicting the distribution of idle resources within the cluster for future job scheduling, as the servers with greater idle resources have more capacity to handle increased resource demands. To reduce inference, optimization objectives such as resource fairness and load balancing are taken into consideration in the works of \cite{Prophet} \cite{Paragon} \cite{Mage} \cite{Hawk} \cite{History-Based}.

Even though there have been many studies on mitigating interference, the complexity and constantly changing nature of large-scale clusters make it difficult to apply benchmark-based and feedback-based solutions. PISM aims to proactively address these challenges by adopting a trace-driven, lightweight, and non-intrusive approach, minimizing server overhead and ensuring a fast response to potential interference.

Despite the extensive research conducted on mitigating interference, the intricate and ever-evolving nature of large-scale production clusters poses challenges in implementing benchmark-based and feedback-based solutions. To address these challenges, PISM adopts a trace-driven, lightweight, and non-intrusive approach to quantity the interference and proactively schedule BEJs to minimize server overhead for potential interference.

\begin{figure*}
    \begin{minipage}[tbp]{0.28\linewidth}
        \centering
        \includegraphics[width=0.9\linewidth]{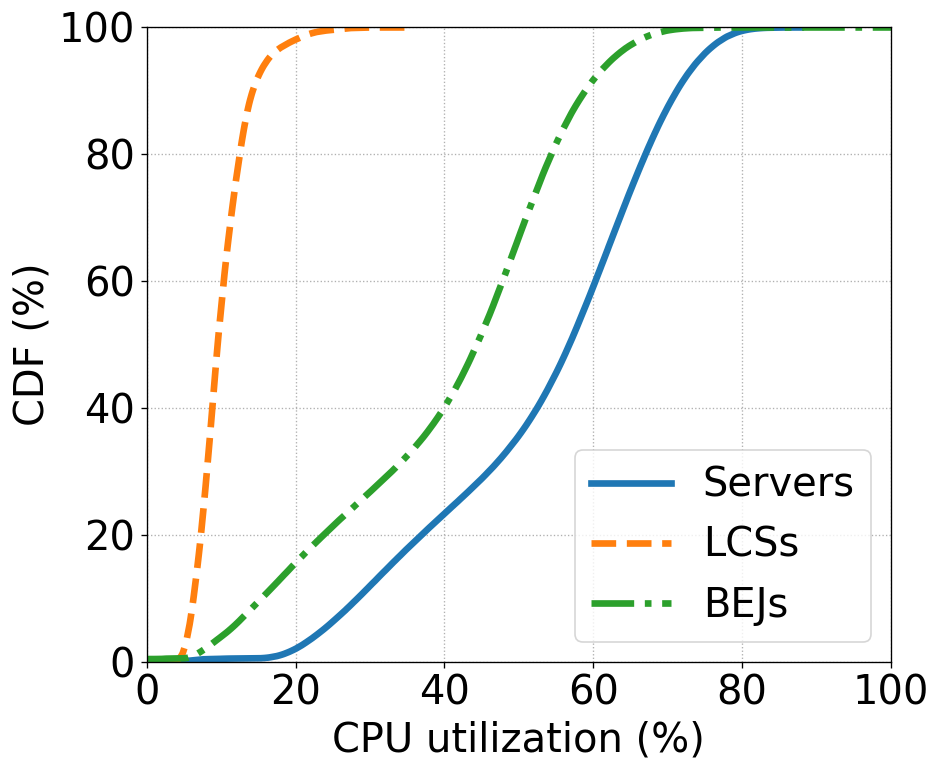} 
        \caption{CDF of CPU utilization of servers, LCSs, and BEJs.} 
        \label{ob1} 
    \end{minipage}
    \noindent
    \hfill
    \begin{minipage}[tbp]{0.28\linewidth}
        \centering
        \includegraphics[width=0.9\linewidth]{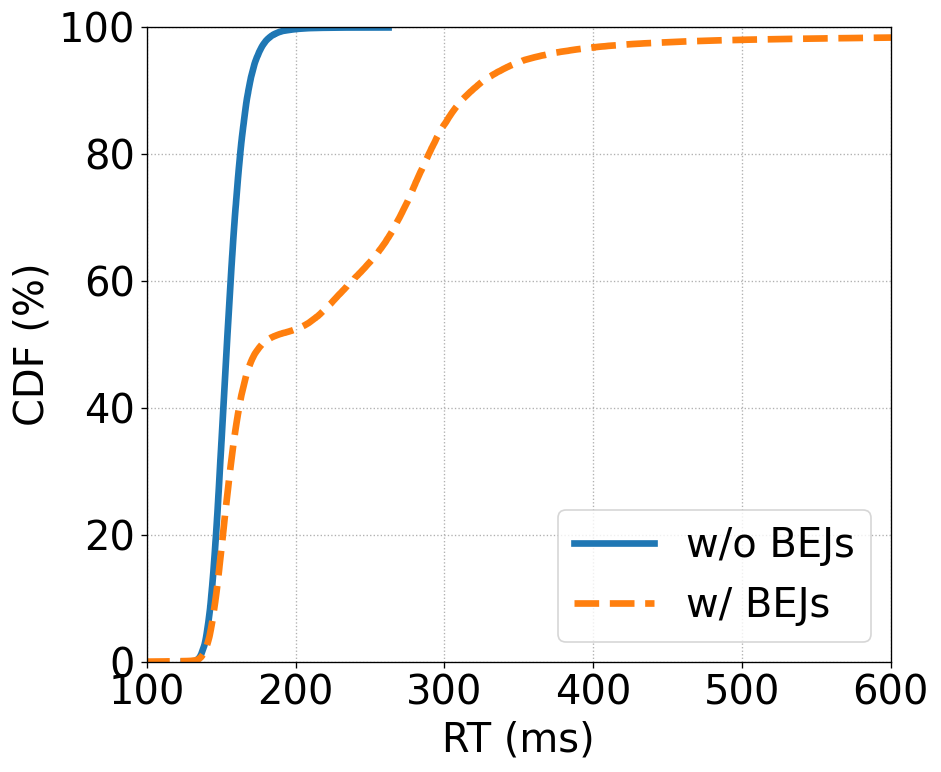} 
        \caption{CDF of \texttt{checkout} instances' RT with and without BEJs.} 
        \label{ob3} 
    \end{minipage}
    \noindent
    \hfill
    \begin{minipage}[tbp]{0.4\linewidth}
        \centering
        \includegraphics[width=1.0\linewidth]{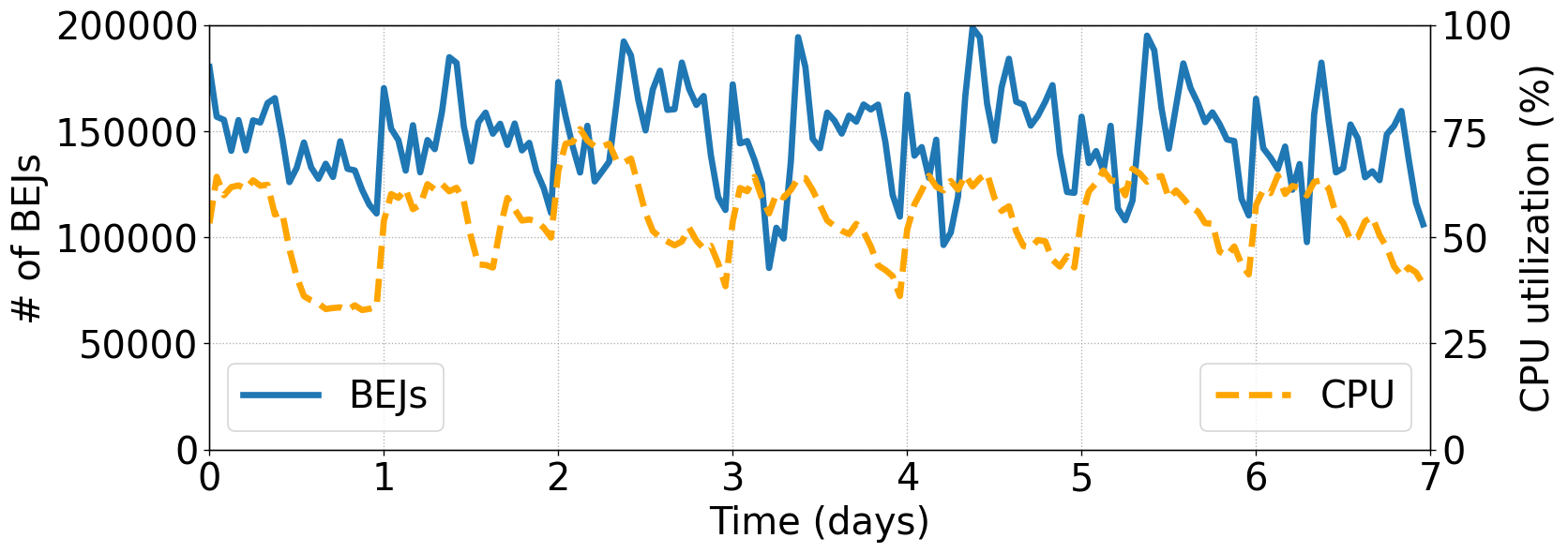}
        \caption{The number of submitted BEJs per hour for the cluster and the average CPU utilization.}
        \label{ov}
    \end{minipage}  
\end{figure*}

\section{Observations}
\label{sec_ob}

We obtained four observations based on the datasets from two large-scale production clusters (Alibaba \cite{MicroserviceCharacterizing} and Google \cite{borg20}), which both co-locate LCSs and BEJs. The statistics on these two datasets are provided in Table \ref{LCBE}. Since LCSs are usually computationally sensitive, we mainly analyze LCSs in terms of resource utilization such as CPU utilization, and application metrics such as response time.

\begin{table}[tbp]
\caption{Statistics of LCSs and BEJs co-located in two production clusters}
\centering
\begin{tabular}{@{}lll@{}}
\toprule
     & Our dataset   &  Google 2019\cite{borg20} \\                           
\midrule
Number of servers & 10,000  & 12,000\\
\midrule
Number of LCSs  & \textasciitilde 25,000           & \textasciitilde   120,000 \\
LCS instances/server & \textasciitilde 20            & \textasciitilde   150 \\
Lifetime of LCS instances      & long-running       & long-running          \\ 
\midrule
Number of BEJs/day & \textasciitilde 4 million     & \textasciitilde 12 million\\
BE instances/server/minute & \textasciitilde 60            & \textasciitilde 66\\
Lifetime of BE instances & 85\% $<$ 1 minute     & 80\% $<$ 15 minutes\\
\bottomrule
\end{tabular}
\label{LCBE}
\end{table}

\noindent\textbf{Observation 1: BEJs dominate the resource usage of servers and are the source of LCS performance interference.}

Figure \ref{ob1} shows the cumulative distribution function (CDF) of CPU utilization of servers, LCSs and BEJs in our cluster. First, although most LCS instances request multiple CPU cores to guarantee the QoS, their utilization is usually below 20\% and has low variance, resulting in resource under-utilization. Second, the CPU utilization of physical servers is almost between 20\% and 80\%. Server-level resource consumption is mainly dominated by BEJs which are also the main cause of changes in server CPU utilization. 

The same pattern exists in other production clusters \cite{borg20} \cite{ResourceCentral}. LCSs in Google clusters have a higher resource usage proportion than ours, accounting for about 20\%-30\% of CPU resources. In addition, BEJs in these clusters are also the main contributors to the overall improvement in resource utilization. 

We find that the performance of LCSs is sensitive to the utilization of BEJs. If we strictly limit the CPU quota of co-located BEJs on the servers, the performance of LCSs is much more stable. However, when BEJs are not restricted with a limited quota, the performance of LCSs fluctuates greatly. 
For example, \texttt{checkout} is one of our most important services that provides the $ordering$ function for online shopping. This service contains more than 3500 instances and has almost 28000 CPU logical cores.
Note that, although we can guarantee the LCSs by limiting the CPU quota of BEJs locally, this method cannot improve the utilization of the cluster.

We also calculate the average RT of all \texttt{checkout} instances with and without the suppression of BEJs. The CDF of \texttt{checkout} RTs in Figure \ref{ob3} shows the sensitivity of \texttt{checkout} on BEJs. With co-located BEJs, the RT of \texttt{checkout} instances varies from 136 to over 1000 ms.



\noindent\textbf{Observation 2: BEJs have high repeatability.} 

We pull the 7-day BEJs executed by a specific business group in our cluster. We analyze repeated BEJs, and the results are listed in Table \ref{ob2}. 
If BEJs have the same DAG, we infer that they are running repeatedly. This business group submitted $282,194$ BEJs in 7 days. Of these jobs, there are almost $1,456$ different DAGs, indicating an average of 200 identical or similar BEJs per DAG. Only 35 BEJs had unique DAGs.
This scenario of repeated BEJs is also reported in the production clusters \cite{Unearthing} \cite{MLaaS}  \cite{Morpheus}.

\begin{table}[tb]
\caption{Statistics of 7-day BEJs of one business unit}
\centering
\begin{tabular}{@{}ll@{}}
\toprule
Description                                 & Count  \\ \midrule
BEJs                                        & 282,194 \\
DAGs                      & 1,456   \\
Infrequent jobs ($<$ 7 jobs with the same DAG) & 1,543   \\
Unique jobs (unique DAG)            & 35    \\
\bottomrule
\end{tabular}
\label{ob2}
\end{table}

Figure \ref{ov} shows the total number of BEJs submitted per hour in our cluster over the 7 days and the hourly average CPU utilization of the entire cluster. We find that some resource-intensive cron jobs are repeatedly executed at midnight and there are fewer requests for LCSs at this time, which leads to the highest CPU utilization of the servers.


\noindent\textbf{Observation 3: LCSs have different interference sensitivities.}

Besides the varying inference levels between LCSs and BEJs, we have observed that LCSs exhibit different sensitivities even when running similar BEJs.
This discrepancy arises from differences in functional logic and resource requirements, which manifest in distinct micro-architecture characteristics, such as CPI (cycles per instruction) and cache misses.

\begin{figure}
    \begin{minipage}[htbp]{0.48\linewidth}
        \centering
        \includegraphics[width=0.85\linewidth]{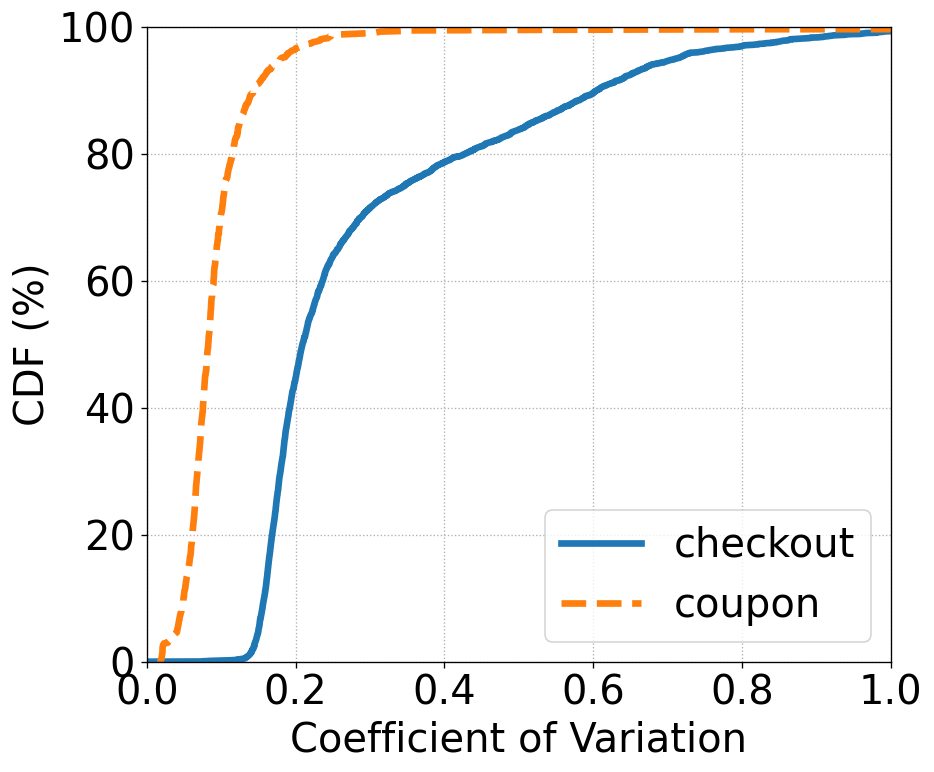} 
        \caption{CDF of coefficient of variation for two different LCSs.} 
        \label{ob4} 
    \end{minipage}
    \noindent
    \hfill
    \begin{minipage}[htbp]{0.48\linewidth}
    \centering
    \includegraphics[width=0.85\linewidth]{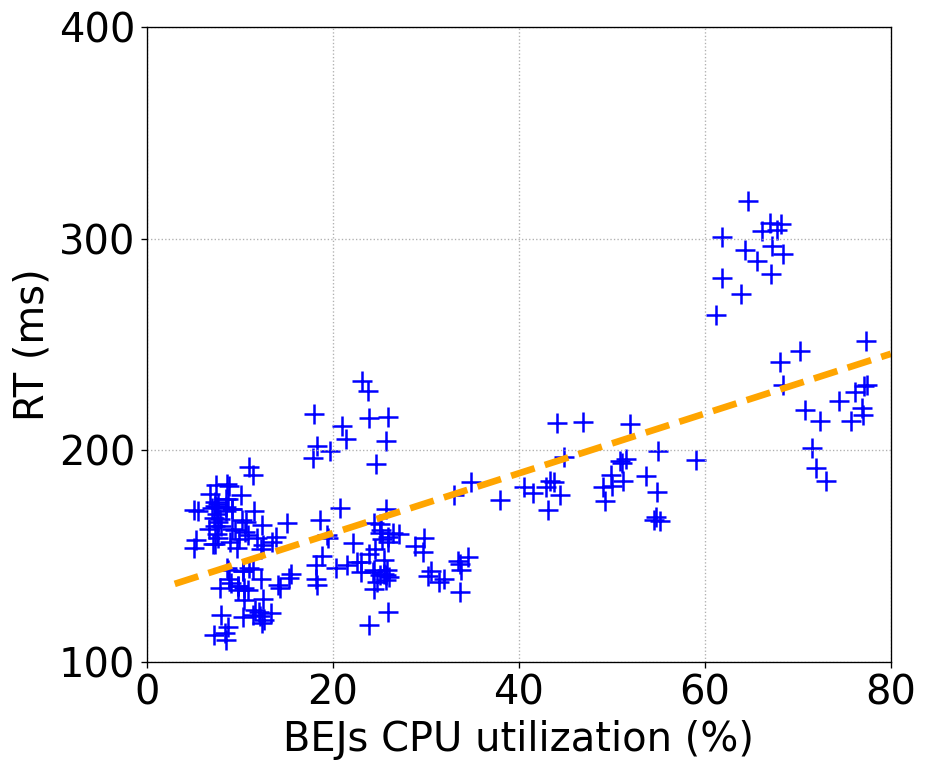} 
    \caption{RT of a \texttt{checkout} instance with different BEJ composition.} 
    \label{ob5} 
    \end{minipage}
\end{figure}

We have calculated the coefficient of variation (CV), which is the ratio of the standard deviation to the mean, for two representative services - \texttt{checkout} and \texttt{coupon} - in order to validate the varying sensitivities of LCSs to interference. The \texttt{coupon} service, responsible for product discounts, consists of 1800 instances and 20000 CPU logical cores. Figure \ref{ob4} illustrates the cumulative distribution function (CDF) of the CV for these two services. The CV of the \texttt{coupon} service remains below 0.2, indicating a relatively stable performance with average response times ranging from 7ms to 9ms. In contrast, the checkout service exhibits much higher volatility, suggesting a more erratic and unpredictable performance when influenced by BEJs. This heightened sensitivity to interference makes the \texttt{checkout} service more susceptible to experiencing long tails, which hampers its quality of service (QoS).

\noindent\textbf{Observation 4: When different BEJ compositions occupy the same amount of CPU on the server, they may cause varying degrees of interference.}

We collected a 3-hour RT trace of a \texttt{checkout} instance and the corresponding total CPU utilization of BEJs running on the same server, as shown in Figure \ref{ob5}. We ensured that all factors other than BEJs remained constant on this server during this period. We find that the RT of the \texttt{checkout} instance varies greatly even if the total CPU usage of BEJs is the same. This indicates that interference is not only affected by the resource usage of BEJs, but is also related to the specific composition of BEJs.


\noindent\textbf{Motivation:} Since BEJs consume most resources and change frequently, it is feasible to evaluate the impact of BEJs on the performance of sensitive LCSs from the perspective of BEJ compositions. In addition, due to the high repeatability of BEJs, the characteristics of new BEJs can be predicted based on historical data, laying a foundation for interference-aware BEJ scheduling. These patterns implied in highly dynamic production clusters inspire us to design PISM.

\section{Design of PISM}
\label{sec_system}

\subsection{Overall Architecture}

When designing PISM, we consider three principles. First, PISM should be lightweight, alleviating the burden on monitoring modules and BEJs schedulers in large-scale clusters. Second, PISM is non-invasive, avoiding the deployment of agents on servers. Third, it should be easy to integrate into any scheduler through an interference scoring scheme, such as Kubernetes.

\begin{figure}[tb]
    \centering
    \includegraphics[width=1\linewidth]{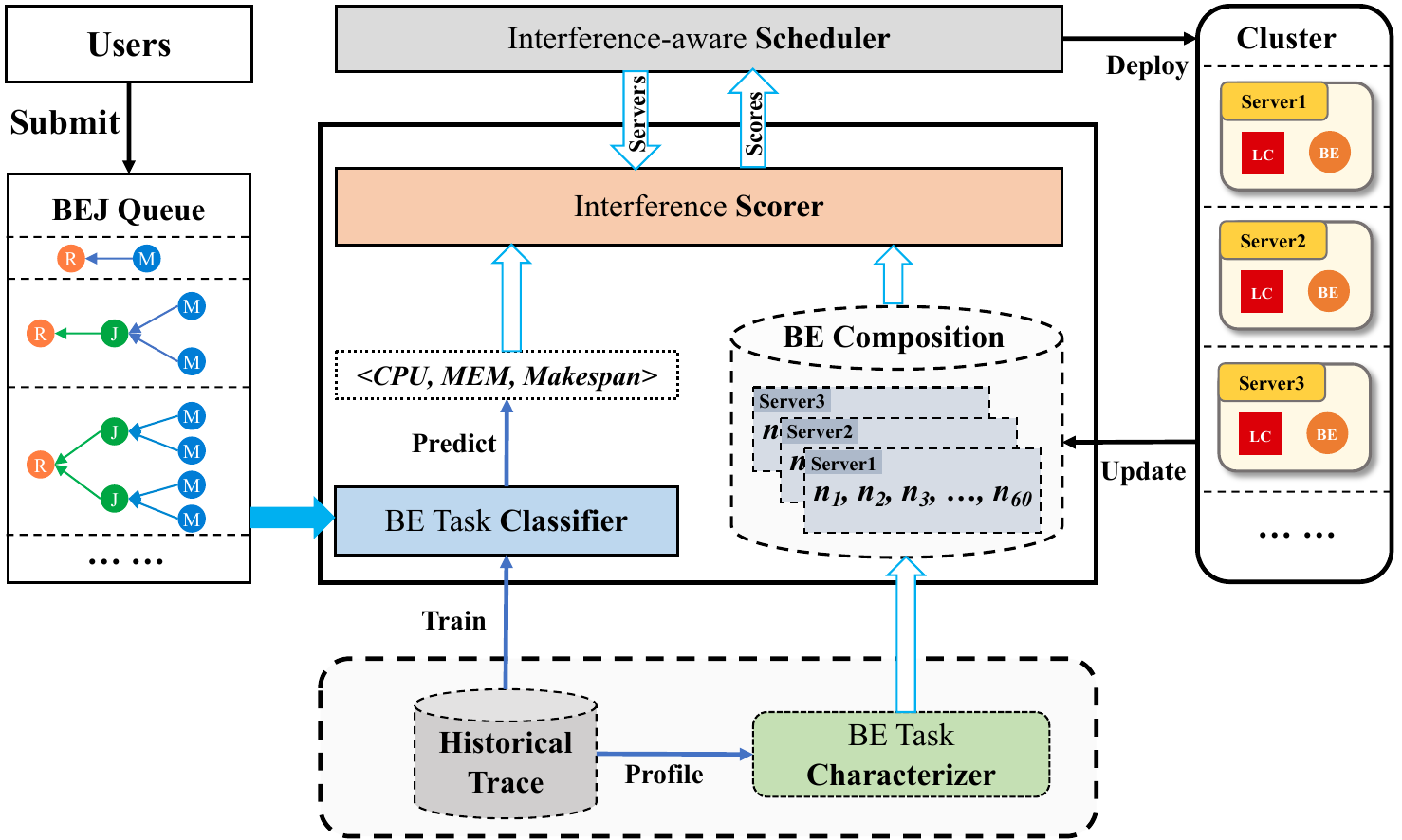} 
    \caption{System architecture of PISM.} 
    \label{system} 
\end{figure}

Figure \ref{system} depicts the system architecture of PISM, which consists of three modules: a BE task characterizer, a BE task classifier, and an interference scorer. Additionally, the system includes two data sets. Historical Trace is utilized to train the task characterizer and BE Composition stores information on the composition of BE instances on each server. The BE Composition update will only be triggered when a new BE instance is scheduled or an existing instance ends.


\begin{figure*}
    \begin{minipage}[tbp]{0.75\linewidth}
        \centering
        \subfigure[CPU\label{CPU}]{
        \includegraphics[width=0.3\linewidth]{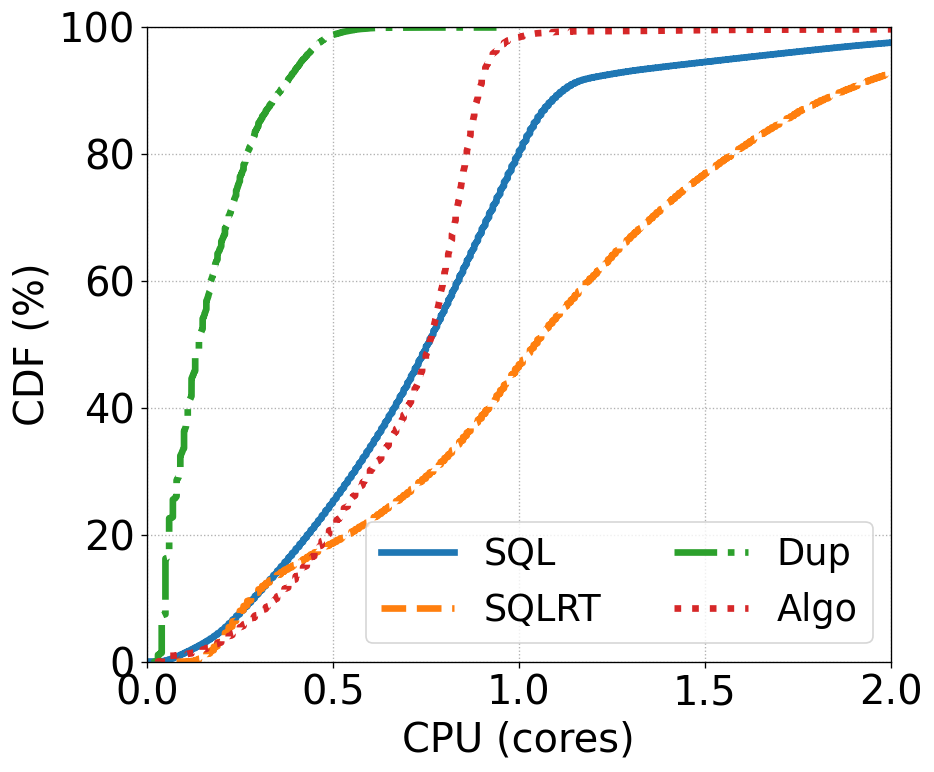}
        }
        \subfigure[MEM\label{MEM}]{
        \includegraphics[width=0.3\linewidth]{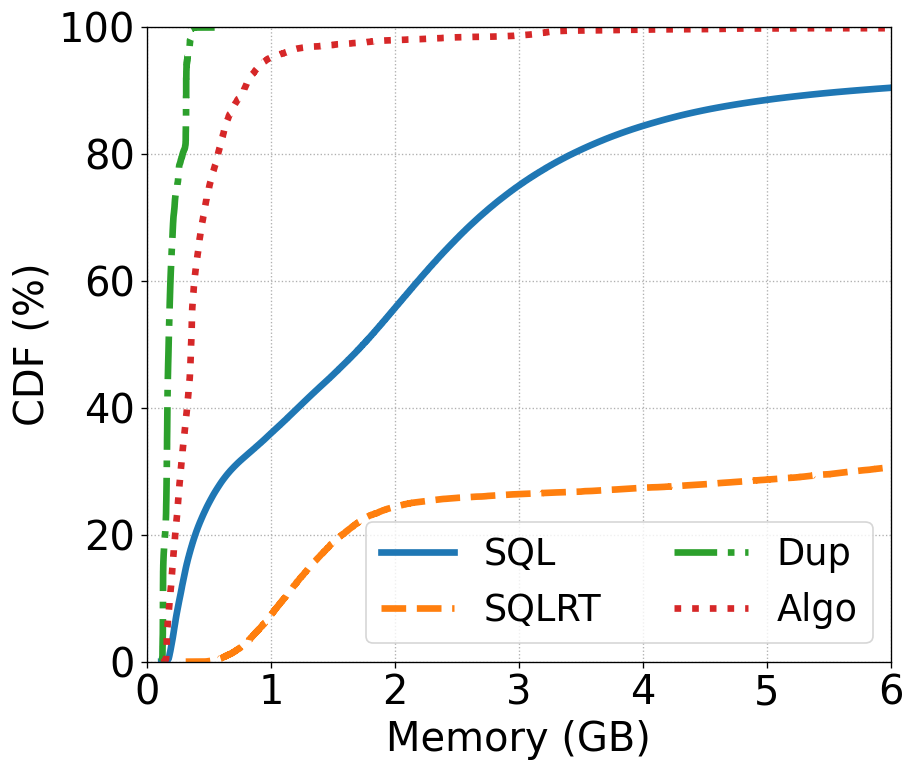}
        }
        \subfigure[Makespan\label{Duration}]{
        \includegraphics[width=0.3\linewidth]{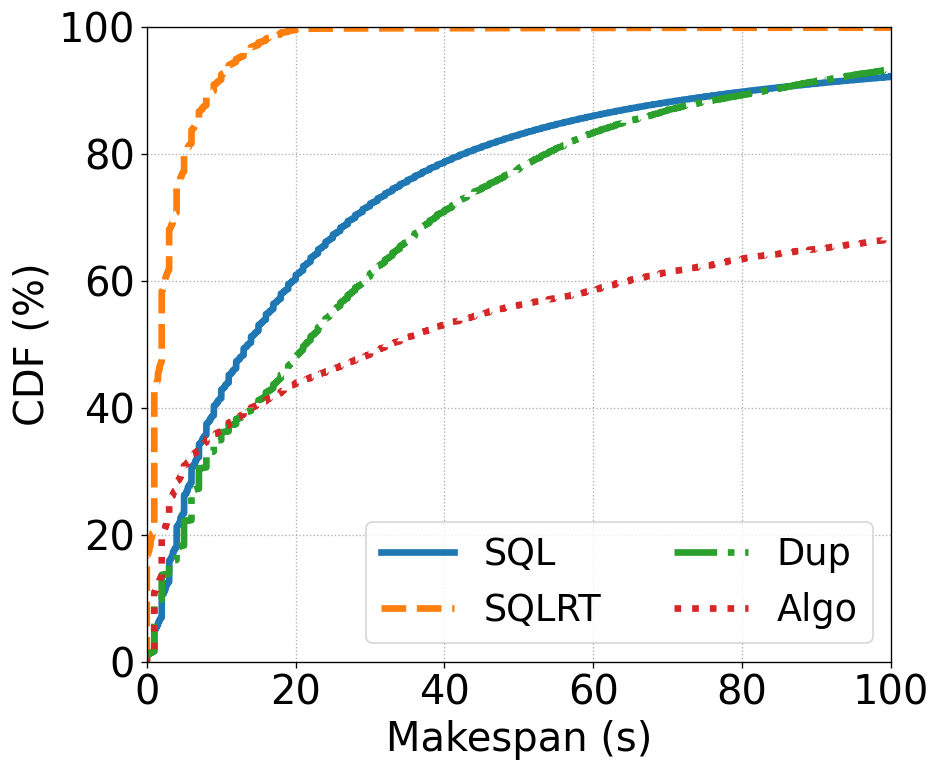}
        }
        \caption{CDFs of CPU, MEM, and makespan of four representative types of BEJs in our production cluster.} 
        \label{tasktype} 
    \end{minipage}
    \begin{minipage}[tbp]{0.24\linewidth}
        \centering
        \includegraphics[width=0.95\linewidth]{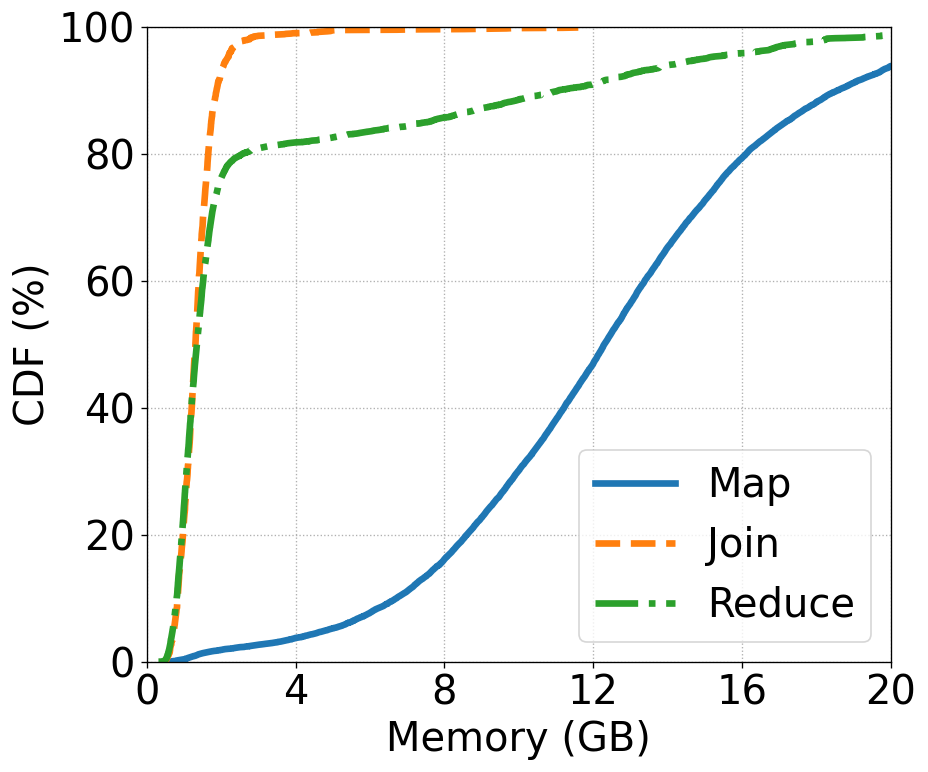} 
        \caption{CDF of memory usage of SQLRT tasks.} 
        \label{SQLRT} 
    \end{minipage}
\end{figure*}

The workflow of PISM is described as follows: First, submitted BEJs are put into the \textbf{Queue}, waiting to be scheduled. Then, when a BE task is ready, PISM first employs the \textbf{BE Task Classifier} to infer its category. Next, given the task category, the \textbf{Interference Scorer} predicts the interference of LCSs and servers if the BE task is deployed on a candidate server. Finally, the \textbf{Interference-aware Scheduler} assigns the BE task to the server with the least interference. Note that PISM is non-intrusive, and can work with any scheduler and be general for other clusters.

\subsection{BE Task Characterization}
\label{sec_character}

Production clusters are highly dynamic due to continuously submitted BEJs. As shown in Figure \ref{ov}, the number of new BEJs per hour is around 100,000 in our cluster. In addition, there are different types of BEJs in production. 
Therefore, it is challenging to study the impact of BEJs on interference from an individual perspective in highly dynamic clusters. Inspired by \textbf{Observation 4}, we explore dividing BEJs into different categories and investigate their collective impact in the form of BE composition.

As mentioned in Section \ref{sec-back}, BEJs are organized in a hierarchical structure: job-task-instance. At the top level, BEJs typically consist of multiple tasks responsible for various operations, such as mapping, joining, or reducing. 
At the lowest level, individual instances of one BE task are executing the same code and handling similar data sizes, which conduct comparable performance and resource usage.
In this way, our characterization of BEJs is primarily designed at the task level. We can effectively capture the overall state and behavior of the BEJs from their corresponding instances.



\subsubsection{Encoding BE Tasks based on Resource Usage}
\label{subsec_profile}

We analyze the task status of four types of BEJs in our cluster for 7 days, namely SQL, SQLRT, Algo and Dup. SQL are typical data processing jobs like Spark; SQLRT jobs require much more resources to ensure that they should be completed as soon as possible; Algo jobs are mainly related to machine learning, such as model training, inference, data normalization, and one-hot encoding; Dup jobs are responsible for data duplication. 

The number of these four categories accounts for 99\% of the total number of BE instances. For each BE task instance, we collect some performance metrics, such as CPU utilization, memory utilization, and makespan, to represent its resource requirements. These metrics are easy to monitor with low overhead.
Meanwhile, the performance metrics of each task are aggregated from all its instances.
We adopt a data-driven approach to classify the BE tasks into different categories and quantify the resource intensity of BE tasks from these performance metrics. For example, SQL jobs are normally CPU-intensive workloads, while Algo jobs are both CPU-intensive and memory-intensive workloads.



\textbf{Encoding based on CPU Usage.} 
To simplify the classification problem, we apply a histogram method to distinguish the intensive of resource usage.
Figure \ref{CPU} demonstrates the CPU usage of BE tasks, which can be divided into three buckets based on their CPU consumption. The workload is traditionally classified as low, medium, or high to represent the degree of CPU consumption. One interesting finding is that the low CPU usage bucket $[0, \alpha_1)$ includes almost all Dup tasks, while the middle CPU usage bucket $[\alpha_1, \alpha_2)$ includes all tasks except for a portion of Algo tasks. The last bucket includes the remaining part of Algo tasks, which are CPU-intensive workloads that consume more CPU than $\alpha_2$. Note that we adopt the settings $\alpha_1=0.5$ and $\alpha_2=1.5$ based on the workload distribution in our experiment.


\textbf{Encoding based on Memory Usage.} As shown in Figure \ref{MEM}, the distribution of memory is more complex compared with CPU resources in our cluster. BE tasks can be divided into five buckets based on four memory thresholds. Specifically, bucket $[0, \beta _1)$ contains all Dup tasks and includes all Algo tasks combining with buckets $[\beta_1, \beta_2)$. Next, bucket $[\beta_2, \beta_3)$ and bucket $[\beta_3, \beta_4)$ contain parts of SQL tasks and SQLRT tasks respectively. Finally, more than 90\% of SQLRT's Map tasks in Figure \ref{SQLRT} require more than $\beta_4$ GB memory, which is typically memory-intensive. We find similar patterns in SQL tasks. In this way, the last bucket we set includes this part of map tasks whose memory usage is greater than $\beta_4$.


\textbf{Encoding based on Makespan.} As for the makespan shown in Figure \ref{Duration}, BE tasks can be divided into four buckets based on three thresholds. Specifically, SQLRT tasks take the least time to complete and can be put into bucket $[0, \gamma_1)$. The makespan of Algo tasks vary widely. Inference or prediction tasks make up a large portion of Algo tasks, and tasks related to model training need to run for tens of minutes or more. So we set the last bucket to include part of the Algo tasks whose makespan is greater than $\gamma_3$. SQL and Dup tasks have similar CDFs, and most of these tasks are completed within $\gamma_3$ seconds. So bucket $[\gamma_1, \gamma_2)$ and bucket $[\gamma_2, \gamma_3)$ contain parts of SQL tasks and Dup tasks respectively.

\subsubsection{Categorization Scheme for BE Tasks}
\label{subsec_category}
Based on the characterization of BE tasks on the three dimensions, the category of BE tasks can be represented by a \textit{triple}:
$$
\setlength{\abovedisplayskip}{4pt}
\setlength{\belowdisplayskip}{4pt}
[CPU, MEM, Makespan]
$$
Specifically, \textit{CPU}, \textit{MEM} and \textit{makespan} have 3, 5, and 4 classes respectively, as described above. So we get 60 BE task categories in total. We essentially categorize BE tasks based on their resource intensity. 
Note that this categorization scheme is not uniform, as certain categories of tasks (e.g., [1, 0, 0]) may account for the majority. This does not affect interference analysis from the perspective of BE composition. 

\subsubsection{BE Composition on Servers}
\label{subsec_BED}
Due to the high dynamics of production clusters, we study the impact of each task category on interference, rather than the effect of one task. All BE task instances co-located on the server can be expressed in the form of BE composition as:
$$
\setlength{\abovedisplayskip}{4pt}
\setlength{\belowdisplayskip}{4pt}
[n_1, n_2, ..., n_{60}]
$$
where $n_i$ represents the number of BE instances of category $i$ on the server. 

\subsection{Classification of New BEJs}
\label{sec_predict}


The high repeatability of BEJs revealed by \textbf{Observation 2} lays the foundation for achieving proactive interference mitigation. Before scheduling, we aim to predict the performance of a new BEJ to determine its resource sensitivity. To do so, we design a hierarchical classification method that first clusters BEJs at the job level according to DAGs and then classifies BE tasks based on the task characterization scheme.

\subsubsection{BEJs Similarity Clustering}
\label{sec_dagsimilarity}

We compute the similarity of BEJs based on their DAG structures. A Graph Kernel is proposed to determine whether two DAGs are isomorphic. The core of the Graph Kernel is to map a graph structure to a vector in the Hilbert space and measure the similarity of two graphs by calculating the inner product of the two vectors. The Graph Kernel has high computational efficiency and is suitable for large-scale clusters that require frequent judgments.

We first cluster BEJs into different groups by computing the inner product matrix of BEJs. Note that each group corresponds to a DAG structure.
After clustering, the inner product between any two BEJs in each group is exactly the same. We use this value as the \textbf{tag} of groups. 
Then, when a new BEJ arrives, we calculate the inner product between it and each of the existing groups (by randomly sampling a BEJ from each group). The new BEJ belongs to the group with a \textbf{tag} equal to the computed inner product. If there is no such group, the new BEJ itself represents a new group. 


\subsubsection{BE Task Classifier}
\label{sec_categorizer}
We combine the DAG \textbf{tag} and some job-level meta-data (e.g., submission time and instance number) as the features of BE tasks. We use these features to train the task classifier which is implemented as an ensemble of three support vector machine classification models based on CPU, MEM, and makespan respectively.

As listed in Table \ref{ob2}, we use six-day data to train the classifier and evaluate its accuracy with one-day data. On average, each DAG structure has about 200 BEJs. We measure classification accuracy on each dimension and overall accuracy (i.e., correct in all three dimensions). 
The results are shown in Figure \ref{categoryp}. At the instance level, the number of instances of BE tasks with accurate classification accounts for about 76\% of the total number of instances, and the accuracy on each dimension reaches 90\%. Based on the evaluation results and Figure \ref{categoryp}, we analyze from two aspects.

First, BE tasks with accurate classification usually have more instances. This is most evident in the MEM dimension, where 60\% of tasks with accurate memory categorization account for 92\% of the total instances. 
Tasks with more instances are more distinctive and stable, which is beneficial for achieving high prediction accuracy. 
Second, the classification error is mainly caused by the boundaries, this problem is more complex and is our future work. 
For example, a task with CPU usage around 0.5 cores may fluctuate between 0.4 and 0.6 cores when running on different servers, thus falling into different categories under the 0.5 threshold. 
Taking into account the boundary error, if we allow the classification for such tasks to fall to the left or right category of the boundary, then the instance-level classification accuracy is up to 99.8\%, 99.8\%, and 99.4\% on CPU, MEM, and makespan, respectively.

\subsection{Interference Scoring Model}
\label{sec_score}
We design an interference scoring model to predict the interference of
LCSs and different BE task compositions on a candidate
server, which will be used in our scheduling model.
Under the same workload, the instance RT of an LCS usually fluctuates within a small range. For example, for \texttt{checkout} with 3500 instances discussed in \textbf{Observation 1}, its 10 percentile (P10) RT is 146ms, and P90 RT is up to 318ms. We find that the higher the instance RT, the greater the interference the instance suffers in our production clusters, which have similar observations in other clusters such as Google Traces \cite{borg20}. Therefore, we use the RT metric to quantify the interference of LCSs and servers.

\subsubsection{Scoring LCS Interference}
In this section, we depict the details of our scoring model with an encoding approach.
We evenly divide the RT of LCS instances into $k$ levels according to the corresponding percentile: the range of $[1,10)$ percentile maps to Level 1 score encoded as [1,0,...,0], the range of $[10,20)$ percentile maps to Level 2 score encoded as [0,1,...,0], and so on. 
In this way, we map the RT of different LCSs to interference scores in the range of $[1, k]$. 
Since the interference level corresponds to RT, it is not only influenced by BEJs, but may also be affected by other factors such as network congestion. When this happens on a server, the interference score of LCS instances on the server will be high, which can effectively prevent BE tasks from being scheduled to the server.

To achieve a balance between performance and accuracy in interference assessment, $k$ can be set to different values in different scenarios. If only interference needs to be detected, $k$ can be set to 2. In this case, the accuracy of the interference assessment can be high, but its value for scheduling is low. Conversely, if larger $k$ values are used to quantify interference with finer granularity, accuracy will decrease.
We set $k = 10$ in the performance evaluations in Section \ref{sec_evaluation}.

\begin{figure}
    \begin{minipage}[htbp]{0.48\linewidth}
        \centering
        \includegraphics[width=0.85\linewidth]{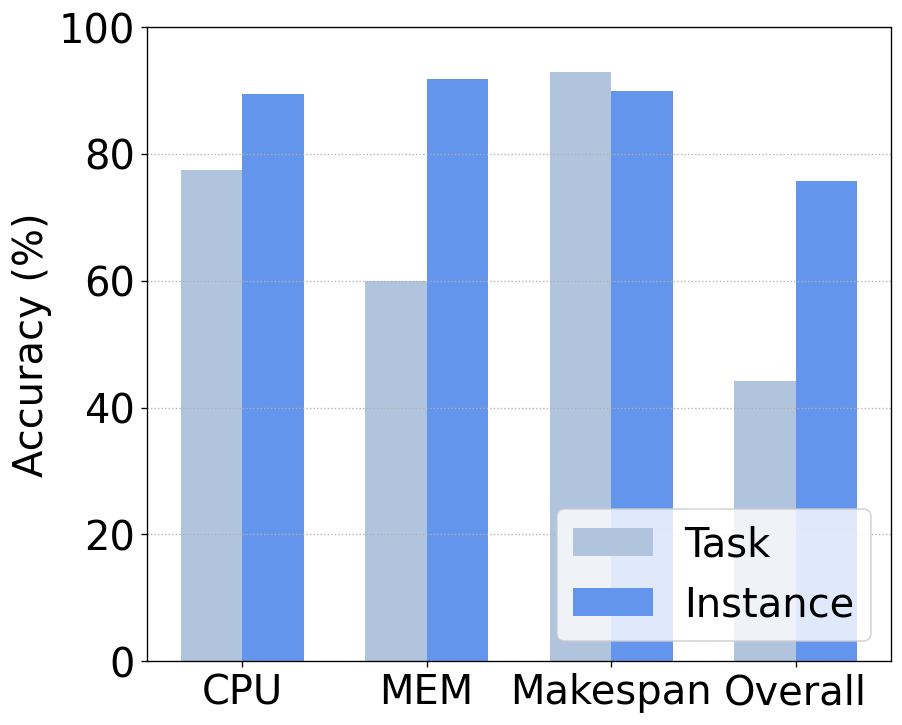} 
        \caption{Classification accuracy at task-level and instance-level.} 
        \label{categoryp} 
    \end{minipage}
    \noindent
    \hfill
    \begin{minipage}[htbp]{0.48\linewidth}
        \centering
        \includegraphics[width=1\linewidth]{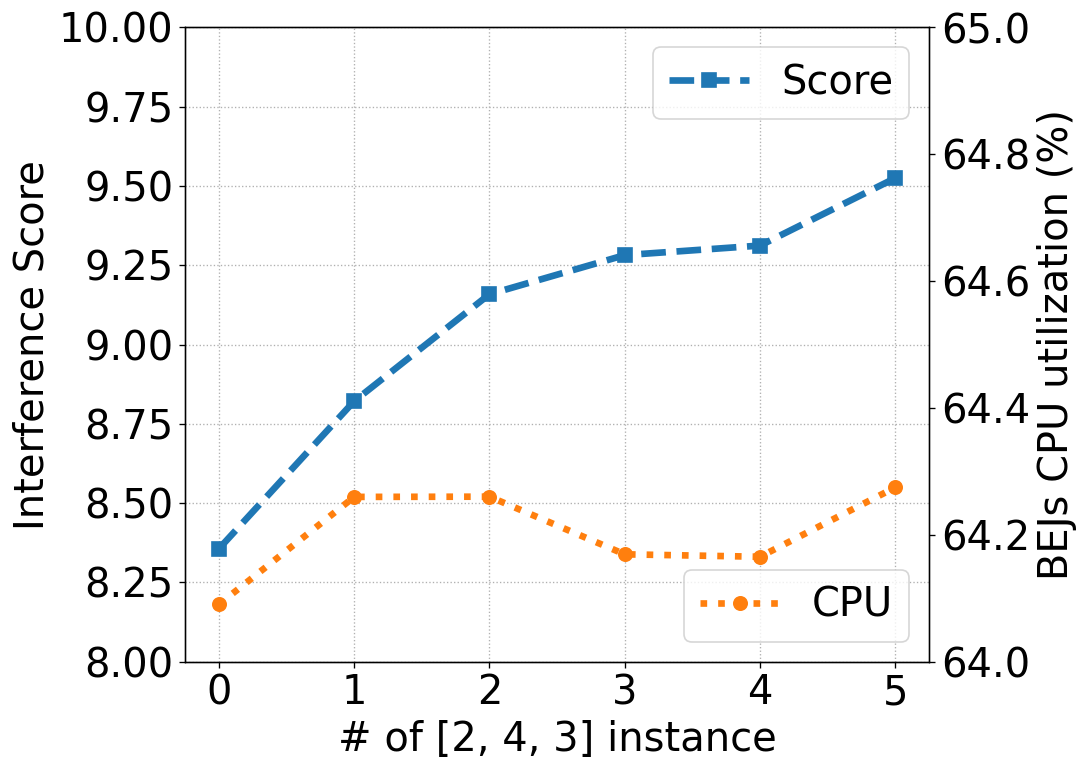} 
        \caption{Interference score of \texttt{checkout} with different number of target instance} 
        \label{case} 
    \end{minipage}
\end{figure}

\subsubsection{Scoring Server Interference}
The level of server interference depends on the LCS instances running on the server. \textbf{Observation 3} in Section \ref{sec_ob} indicates that different LCSs have diverse sensitivity to interference. Based on the LCS interference scoring model, a key issue is how to weight each LCS instance on the server to comprehensively quantify server interference. Given the weight $w_i$ of each LCS instance, the server interference score can be aggregated by:
$$
\setlength{\abovedisplayskip}{4pt}
\setlength{\belowdisplayskip}{4pt}
Score_{server} = \sum w_i * Score_{_i}
$$
where $i$ iterates over all LCS instances on the server, and $Score_i$ is the interference level of the $i-th$ LCS at this server. 


We evaluate 4 weighting schemes in Section \ref{subsec_acc}. (1) Fair: all LCS instances have the same weight; (2) CPU-Quota: the weight of LCS instances is determined by their requested CPUs; (3) Corr: the correlation coefficient between the RT of LCS and the total CPU usage of BE tasks on the server; and (4) CV: the coefficient of variation of LCS instance RTs.



\subsection{Interference Prediction based on Server BE Composition}
\label{subsec_relation}

In this section, we provide a decision tree model to predict LCS interference based on server BE compositions.
\textbf{Observation 4} indicates that LCS interference is affected by the different server BE compositions,  which represents the overall resource competition generated by all co-located BE instances on the server. In general, the more resource-intensive BE tasks co-located, the greater the chance of resource contention between BEJs and LCSs.

We provide an example to analyze how resource-intensive BE tasks interfere with \texttt{checkout} instances. The category of target BE task is [2, 4, 3], which has the largest CPU and memory usage and the longest makespan.
We strictly limit the CPU utilization of all co-located BE instances (about 64\%) and the workload of \texttt{checkout} to ensure as many consistent conditions as possible, except for the different composition of BE instances. Since \texttt{checkout} has 3500 instances, there are enough samples for each number of target instances. 





Given $k = 10$, we calculate the average interference score of \texttt{checkout} with a different number of target instances, as shown in Figure \ref{case}. 
Although the overall CPU usage of BE instances is almost the same, increasing the number of target instances increases the interference score of \texttt{checkout}. 
This indicates that large tasks like category [2, 4, 3] should be distributed across servers, paired with small tasks. 

In summary, for each new BE task, PISM first determines the category representing its resource intensity. Suppose the task is scheduled on a candidate server, PISM predicts the interference level of all co-located LCSs on the server based on server BE composition considering the impact of the new BE task. After evaluating all candidate servers, the scheduler selects the server with the least interference to deploy the task.



\section{Evaluation}
\label{sec_evaluation}

We evaluate the effectiveness of PISM based on real traces from a large-scale production cluster, and we also conduct our scheduling algorithm on a small-scale cluster with 14 servers.


\begin{figure}[tbp]
\centering
\includegraphics[width=0.9\linewidth]{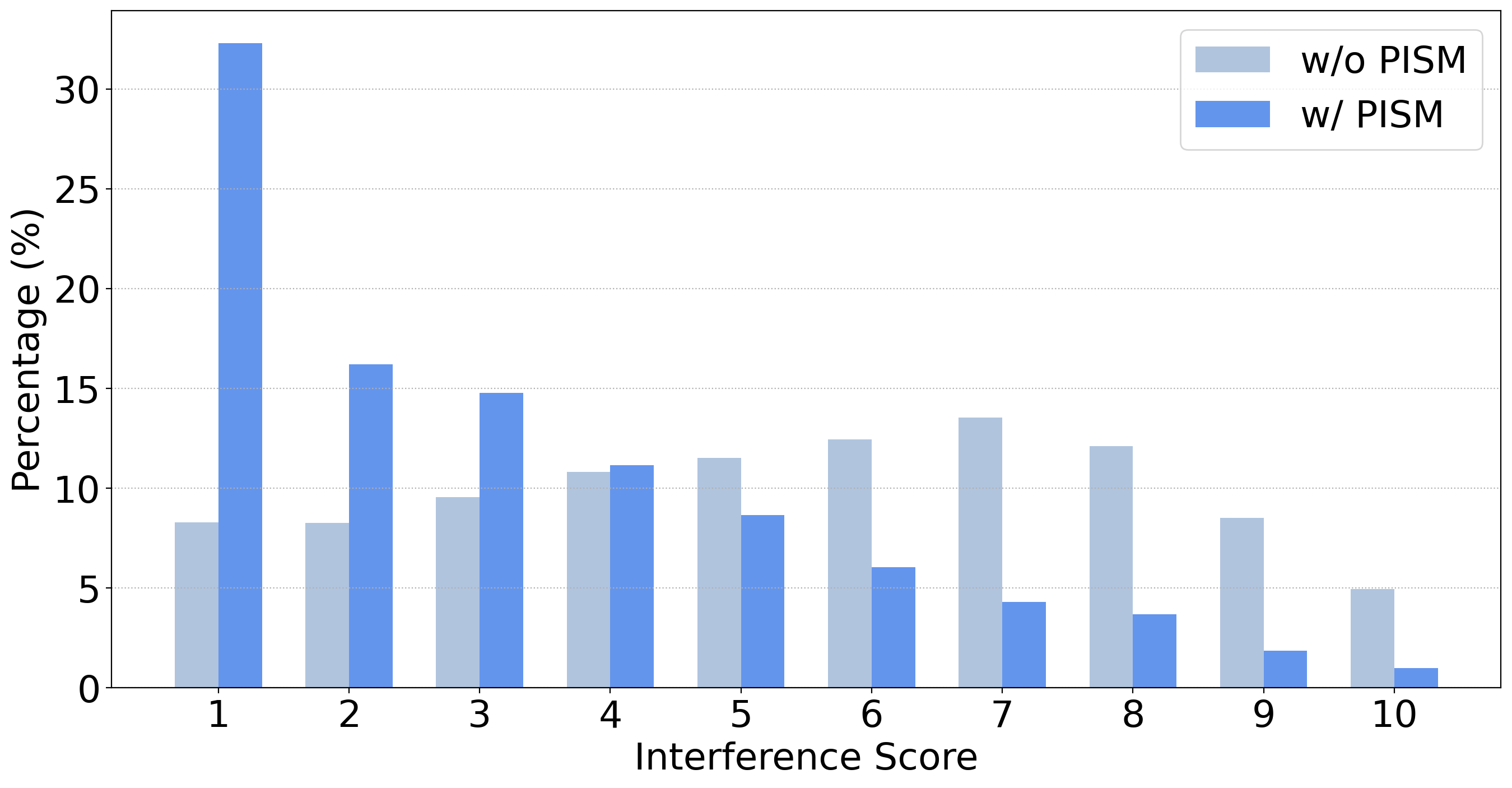} 
\caption{Distribution of server interference scores with and without PISM.} 
\label{cmp10}
\end{figure}

\subsection{Large Trace-based Simulations}
\subsubsection{Setup}
We collect BEJ scheduling traces on 600 servers over 1 week and conduct a simulation study by rescheduling these BE instances to evaluate the effectiveness of PISM. 
The CPU model of the servers is Intel(R) Xeon(R) Platinum {\bf 826X} CPU @ 2.50GHz (Cascade Lake model). The base frequency is 3.2 GHz and the L3 cache size is 36608 KB. Non-uniform memory access (NUMA) is enabled to increase the overall bandwidth and reduce the latency to memory.

The workloads consist of hundreds of LCSs and $282194$ BEJs in 7 days. One important LCS is \textbf{checkout}, having more than 3500 instances and consuming almost 28000 CPU cores. For BEJs, there are almost $1456$
different DAGs, indicating an average of 200 identical or similar BEJs per DAG, which is general and reported in other production
clusters such as Google \cite{googleworkload}.

The simulator contains the \textbf{BE Task Classifier} to infer the categories of tasks and the \textbf{Interference Scorer} to predict the interference score when scheduling, as described in Section \ref{sec_system}. 
After predicting the category of each BE instance, the scheduler selects the candidate server with the lowest interference score for scheduling. 
The optimization goal of PISM is to ensure that the same BE instances are scheduled while reducing the average interference level of all servers. 


\subsubsection{Simulation Results}
Figure \ref{cmp10} shows the simulation results.
In the original trace, there is not much difference in server interference scores, which is the \textbf{Baseline} method (without PISM) and is the default scheduling algorithm in our cluster.
We implement our PISM algorithm by continuously scheduling BE instances to our cluster based on their submission time.
Moreover, PISM reschedules BE instances that are originally assigned to servers with high interference to servers with low interference. 
We calculate the server interference score based on the formula in Section \label{sec_score} and classify the interference degree into ten buckets. 
The score in level-0  denotes the lowest interference, and the score in level-10 has the highest chance for resource contention. 
As a result, PISM increases the workloads of servers with the lowest interference from 8.3\% to 32.3\% in Figure \ref{cmp10}. We also discover that our algorithm has the largest optimization to mitigate interference in level-10, reaching 82.1\%. 
In summary, PISM reduces the server interference score from 5.468 to 3.197, mitigating performance interference by 41.5\%. This means that PISM can effectively alleviate the long-tail phenomenon.

\begin{table}[tbp]
\caption{Throughput improvement of 150 LCSs.}
\centering
\begin{tabular}{@{}lll@{}}
\toprule
           & Percentage & Improvement \\ \midrule
Tail-heavy & 9.3\%      & 76.4\%              \\
Tail-light & 90.7\%     & 17.2\%              \\ \bottomrule
\end{tabular}
\label{throughput}
\end{table}

We also study the response time (RT) and throughput of 150 critical LCSs on 600 servers. To simplify the evaluation, we classify these LCSs into two categories: $tail-heavy$ and $tail-light$. We find that the P90 RT of $tail-heavy$ LCSs is 5x the average RT, and the rest are $tail-light$. Table \ref{throughput} illustrates the percentage of LCSs in two categories and their respective throughput improvements.  $Tail-light$ LCSs make up most of our cluster, with an average throughput increment of 17.2\%. The throughput improvement on $tail-heavy$ LCSs is even more significant, reaching 76.4\%.
This suggests that PISM not only effectively reduces the percentage of P90 RT, but also benefits the LCSs with high tail latency.

\subsubsection{Integration with Schedulers}
We examine the generality and performance of PISM, which can be seamlessly integrated into other schedulers. First, we consider two basic schedulers: \textit{spread} and \textit{stack}, which are also used in Kubernetes\footnote{The scheduling architecture in our cluster is similar to Kubernetes.}, Spark, Docker Swarm, and other frameworks \cite{MLaaS, jobplacement, spread-n-share}. \textit{spread} schedules tasks to the server with the lowest CPU utilization to balance the CPU utilization of the cluster, while \textit{stack} selects the server with CPU utilization as close to the threshold as possible. We use 80\% as the CPU threshold because there are very few servers in our cluster that have more than 80\% CPU utilization, as shown in Figure \ref{ob1}.

\begin{figure}
\centering
        \subfigure[Interference score\label{cmb-score}]{
        \centering
        \includegraphics[width=0.45\linewidth]{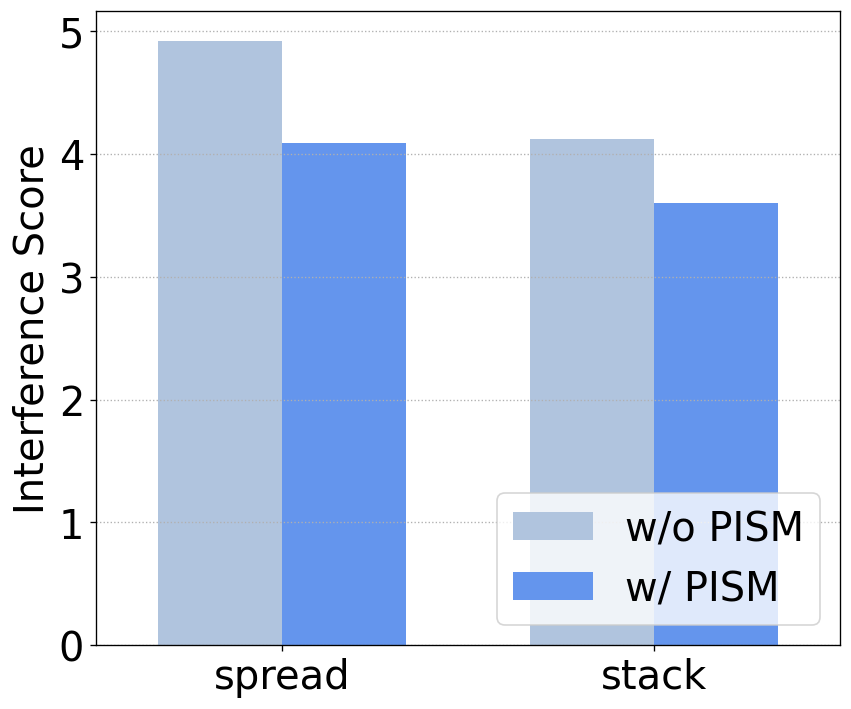}
        }
        \subfigure[CPU utilization\label{cmb-cpu}]{
        \centering
        \includegraphics[width=0.47\linewidth]{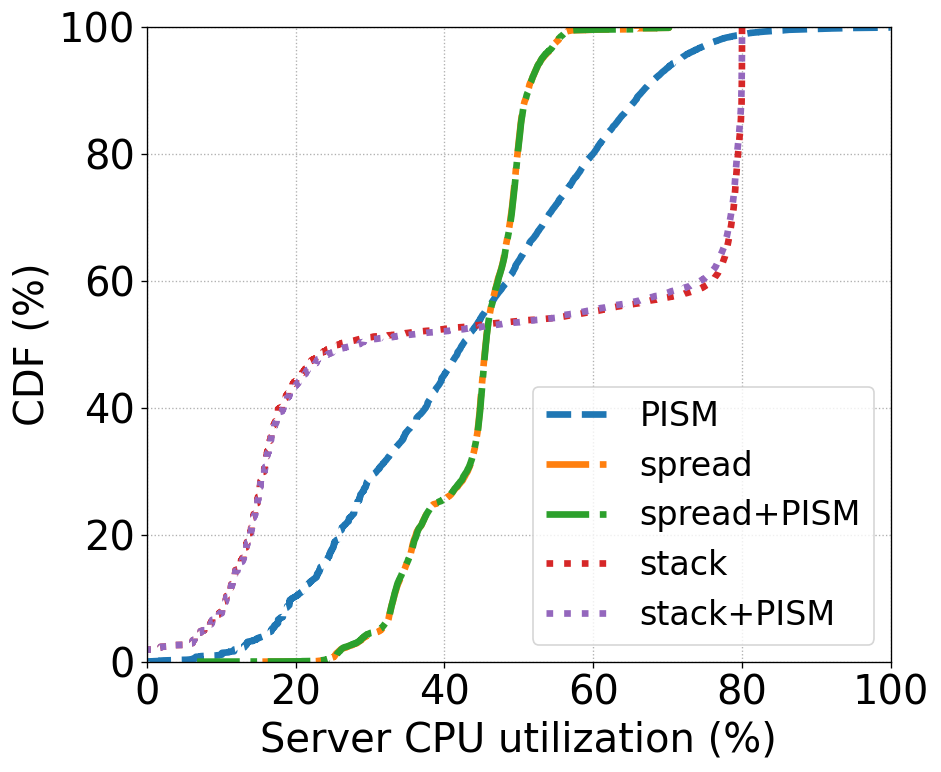}
        }
    \caption{Scheduling using PISM, \textit{spread}, \textit{spread}+PISM, \textit{stack}, \textit{stack}+PISM.} 
    \label{cmb} 
\end{figure}

Next, we employ both schedulers to select 10 candidate servers to replace one server. Subsequently, we predict their interference level with or without PISM, and then schedule task instances to the candidate server with the lowest interference score.
As illustrated in Figure \ref{cmb-score}, PISM-based schedulers can effectively reduce the overall interference by 16.9\% and 12.5\% compared with \textit{spread} and \textit{stack}, respectively.
\textit{spread} focuses on load balancing, resulting in little difference in server CPU utilization, basically distributed between 40\% and 50\%, as shown in Figure \ref{cmb-cpu}. 
On the other hand, \textit{stack} considers resource reservation, leading to polarized results. 
For \textit{stack}, there are many servers with the lowest interference score, about 15\% higher than \textit{spread}, resulting in relatively low overall cluster interference.
Figure \ref{cmb-cpu} also shows that PISM does not change the goals of these two schedulers while significantly reducing interference. 

\begin{figure}[tbp]
\centering
        \subfigure[\texttt{checkout}\label{avg_rt_checkout}]{
        \centering
        \includegraphics[width=0.47\linewidth]{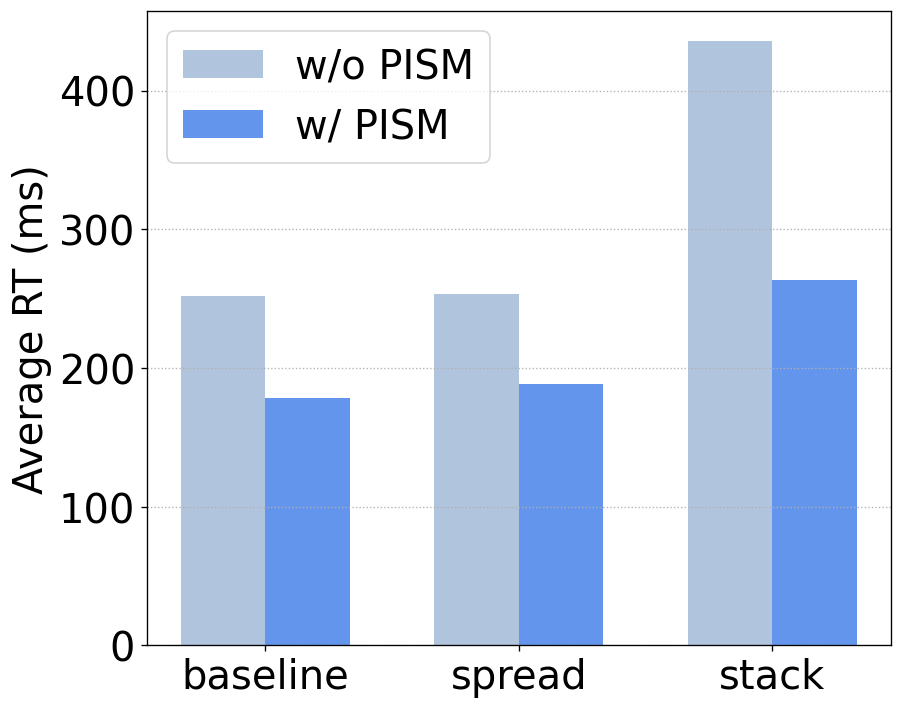}
        }
        \subfigure[\texttt{browse\_product}\label{avg_rt_product}]{
        \centering
        \includegraphics[width=0.47\linewidth]{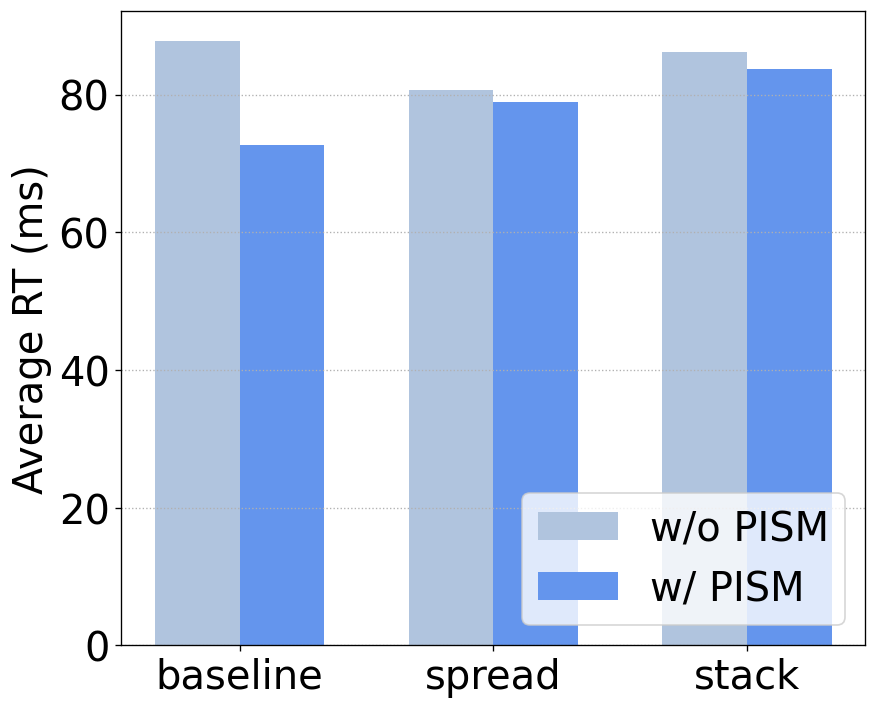}
        }
    \caption{Average RT of \texttt{checkout} and \texttt{browse\_product} with different schedulers.} 
    \label{AverageRT} 
\end{figure}

\subsection{Small-scale Cluster Experiments}
\subsubsection{Setup}
To investigate the impact of PISM on the actual RT of LCSs, we set up a small-scale Kubernetes cluster consisting of 14 servers. Each server is equipped with an Intel(R) Xeon(R) Platinum 8269CY CPU model and 512GB of memory. We use Online Botique\footnote{Google’s open source e-commerce microservice platform, \url{https://github.com/GoogleCloudPlatform/microservices-demo}} and Spark to simulate LCSs and BEJs. Their workloads are generated based on the production cluster trace. PISM is integrated into three schedulers, \textit{spread}, \textit{stack}, and random placement.

\subsubsection{Experiment Results}
Figure \ref{AverageRT} shows the average RT of two critical services of Online Botique, \texttt{checkout} and \texttt{browse\_product}. PISM significantly reduces the RT of the two services with different schedulers. The maximum reduction is achieved when PISM is paired with \textit{stack}, reducing the average RT of \texttt{checkout} from 436ms to 263ms, a decrease of 39.5\%. 
Compared to \texttt{browse\_product}, the RT improvement of \texttt{checkout} is more significant, because \texttt{checkout} has a more obvious long-tail latency. 

\begin{figure}[tbp]
\centering
        \subfigure[\texttt{checkout}\label{PRT_checkout}]{
        \centering
        \includegraphics[width=0.47\linewidth]{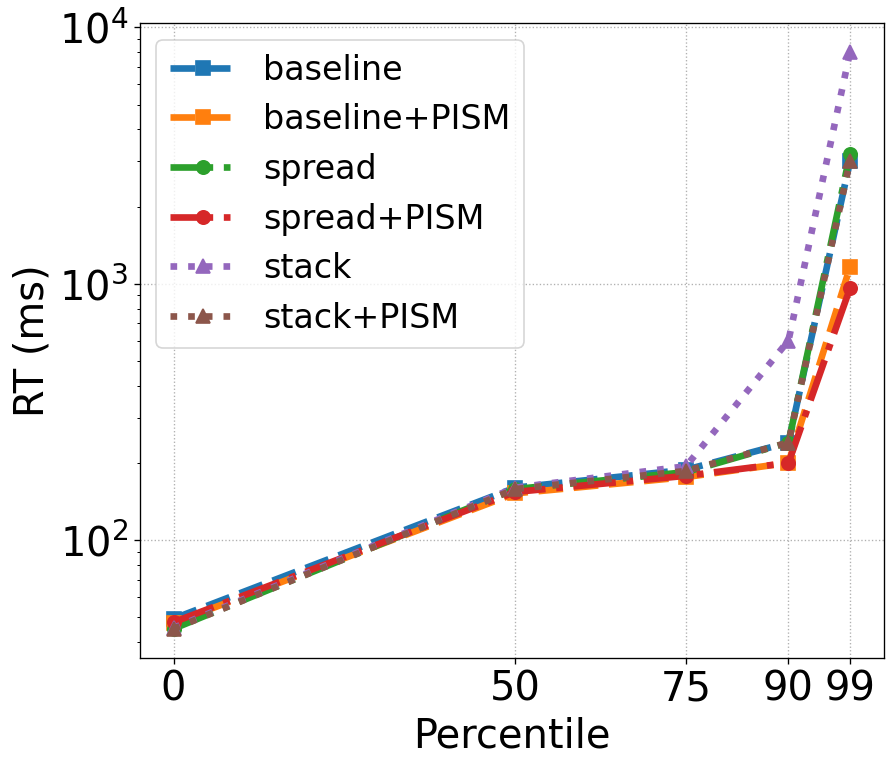}
        }
        \subfigure[\texttt{browse\_product}\label{PRT_product}]{
        \centering
        \includegraphics[width=0.47\linewidth]{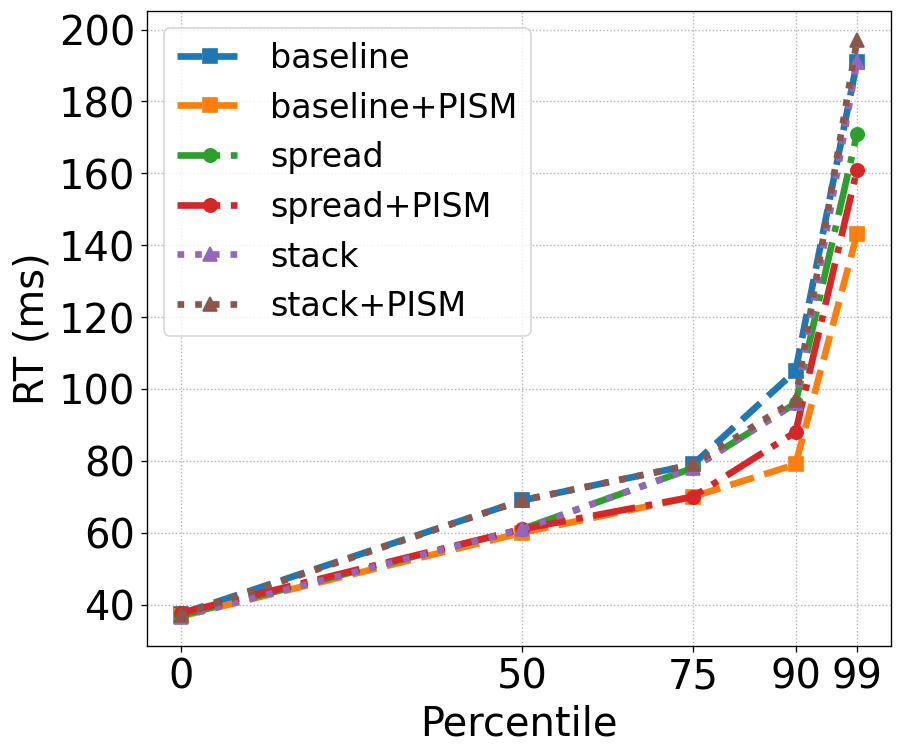}
        }
    \caption{P0, P50, P75, P90 and P99 RT of \texttt{checkout} and \texttt{browse\_product} using different schedulers.} 
    \label{PRT} 
\end{figure}

Figure \ref{PRT} shows five percentile RT values of \texttt{checkout} and \texttt{browse\_product}. For the two services, there is little difference in the P0, P50, and P75 RT with or without PISM. However, there are significant improvements in the P90 and P99 RTs of \texttt{checkout} with PISM, especially for the P99 RT reduced from 8000ms to 3000ms. This demonstrates the effectiveness of PISM in improving service performance by alleviating the long-tail phenomenon. These results are consistent with the simulation results.

\begin{figure}[tbp]
\centering
\includegraphics[width=0.9\linewidth]{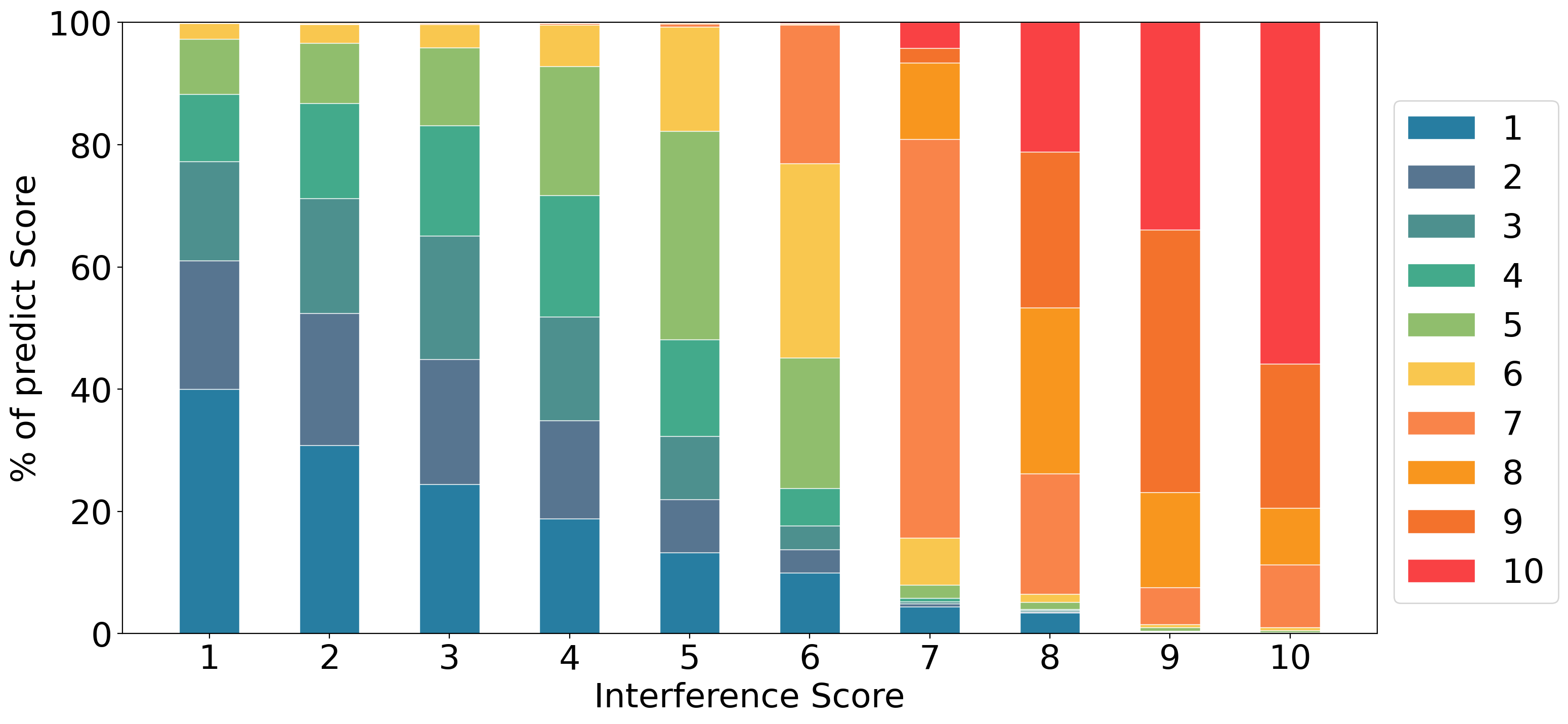} 
\caption{Distribution of predicted interference of \texttt{checkout}.} 
\label{sl} 
\end{figure}

\subsection{Accuracy of Interference Prediction}
\label{subsec_acc}
\subsubsection{Interference Scores of LCSs}
We collect a 2-day of trace from our cluster, one day as training set and another day as ground truth to evaluate the interference prediction model of LCSs. 
As shown in Figure \ref{sl}, we evaluate the interference predictions of \texttt{checkout}. The predictions are centered around each true interference score. Since the interference score represents a trend and needs to be compared with each other when scheduling, we consider predictions to be accurate with an absolute error of no more than 2, calcuted as, $|predicted~score - true~score| \leq 2$. The results demonstrate that we can achieve a prediction accuracy of 87.2\% for \texttt{checkout}.

We further focus on the case of relatively high interference and low interference (i.e., evenly divided into two categories). We treat those with large scores as BUSY, and the opposite as IDLE. This can be viewed as a simple binary classification problem of whether interference occurs. We then compute the precision, recall, and F1-score metrics, as listed in Table \ref{PR}. Our model performs well on this binary classification problem with F1-scores above 0.9. 
The reason for the higher precision but lower recall of the BUSY class is that our model only considers BEJ, ignoring interference caused by other factors (such as other applications). 
The IDLE class is the opposite. Figure \ref{sl} also shows that when the true interference score is greater than 5, there are still a certain number of samples whose level is predicted to be 1. One possible explanation is that the interference might be caused by other LCSs on the same server, which cannot be reflected by BE composition.

\begin{table}[tbp]
\caption{Prediction evaluation of BUSY and IDLE for \texttt{checkout}.}
\centering
\begin{tabular}{@{}llll@{}}
\toprule
     & Precision & Recall & F1-score \\ \midrule
BUSY & 92.5\%    & 87.9\% & 0.902    \\
IDLE & 88.6\%    & 93.0\% & 0.907    \\ \bottomrule
\end{tabular}
\label{PR}
\end{table}

When the real interference level is very high (e.g., score $>$ 8), it is harmful to predict it as a small value (e.g., score $<$ 3). 
In this case, the scheduler will continue to schedule tasks on such a heavy-loaded server, eventually degrading the performance of LCSs. Fortunately, the probability of this happening in our model is only 0.04\%. Conversely, predicting a low level as a high level may result in missed scheduling opportunities, which is only a 0.006\% chance for our model.

\begin{figure*}
    \begin{minipage}[tbp]{0.24\linewidth}
        \centering
        \includegraphics[width=0.95\linewidth]{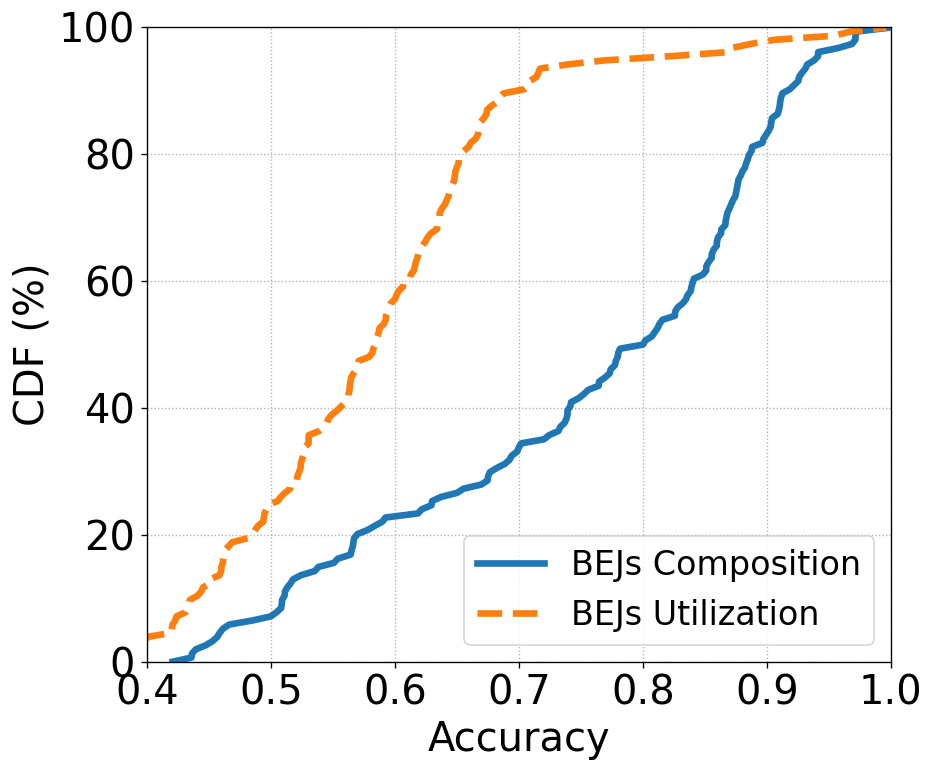} 
        \caption{CDF of interference score prediction accuracy of 150 LCSs.} 
        \label{slall} 
    \end{minipage}
    \noindent
    \hfill
    \begin{minipage}[tbp]{0.75\linewidth}
        \centering
        \subfigure[Accuracy\label{mlacc}]{
        \includegraphics[width=0.3\linewidth]{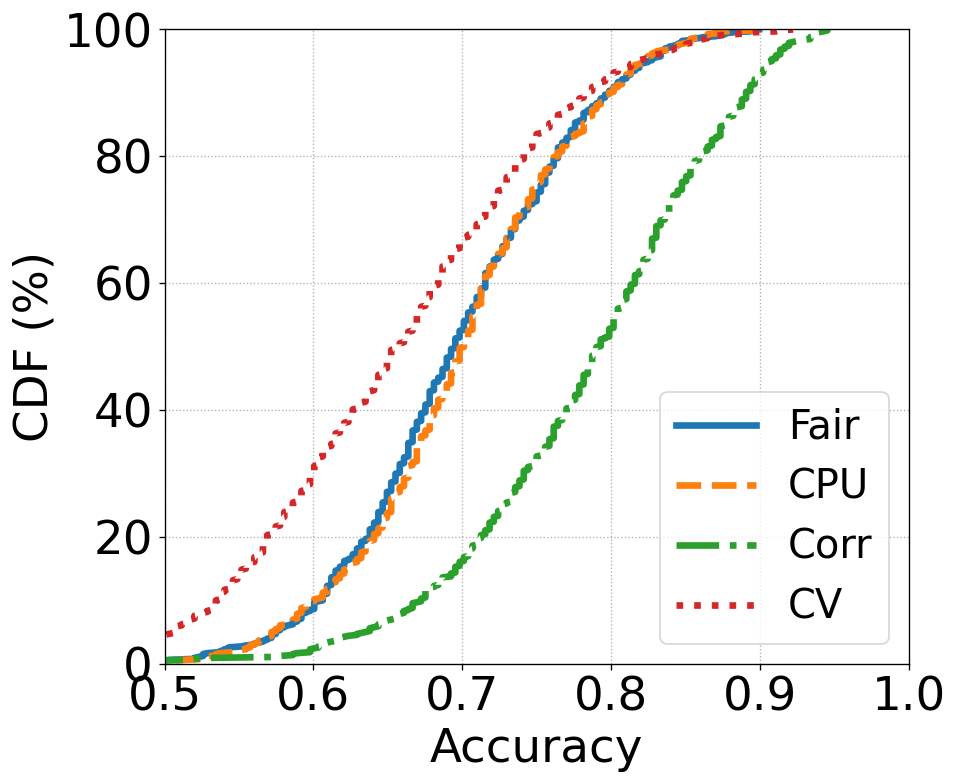}
        }
        \subfigure[F1-score\label{mlf1}]{
        \includegraphics[width=0.3\linewidth]{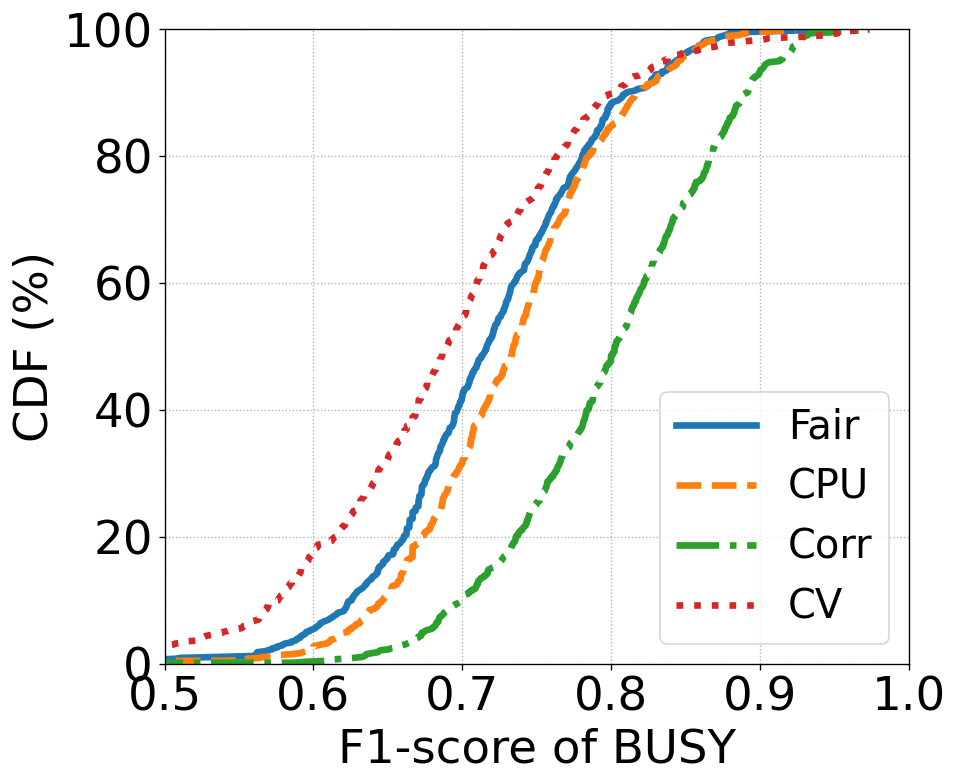}
        }
        \subfigure[Severe errors\label{mlrisk}]{
        \includegraphics[width=0.3\linewidth]{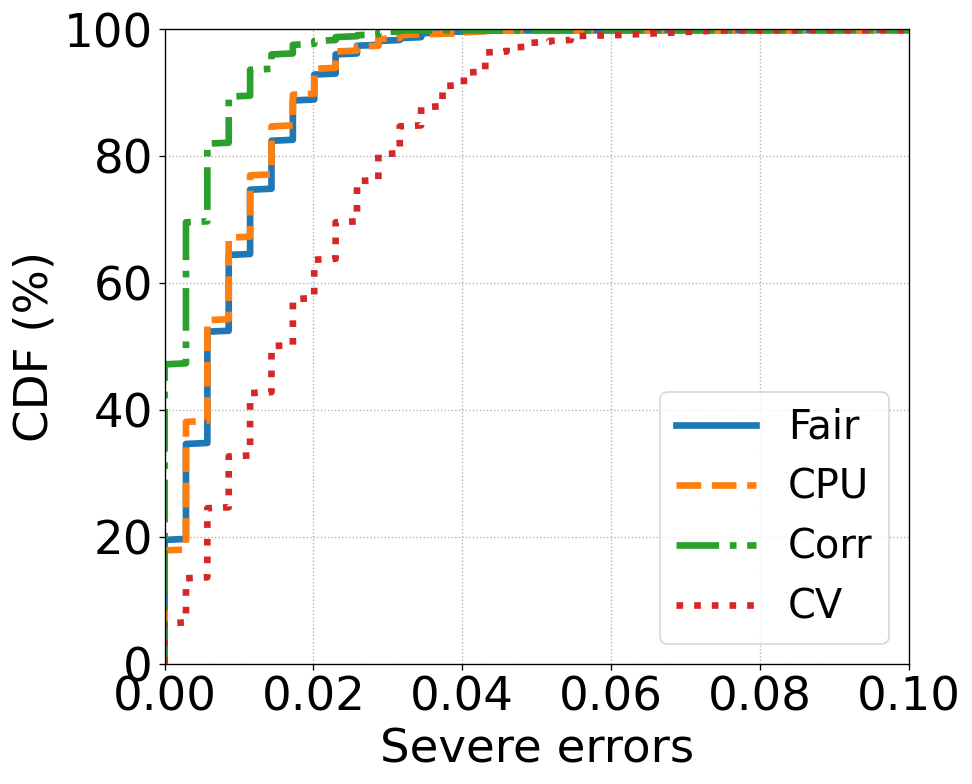}
        }
        \caption{Evaluation of interference predictions for 600 servers.} 
        \label{ml} 
    \end{minipage}
\end{figure*}

We then investigate the accuracy of interference prediction shown in Figure \ref{slall}. Here we consider 150 crucial LCSs in our production cluster, and each one has more than 10 instances.
The baseline method ($BEJs Utilization$) calculates the LCS interference score based on the CPU usage of all LCSs as observation 4 shows there is a linear relationship between the overall BEJ resource usage and LCS RT.
Our method $BEJs Composition$ estimates the interference score according to the BEJs composition.
Figure \ref{slall} shows the average accuracy of the BEJ utilization-based method is only 58.3\%, while the BEJ composition-based method is 75.2\%. 
This verifies that the BE task characterization method captures performance interference patterns. 
Notably, the BEJ composition-based method itself is simple and efficient, laying a good foundation for deployment in large-scale clusters.
In contrast, if we want to predict the interference score of LCSs based on BEJ utilization, we need to collect the resource utilization of each server in real-time and predict the exact run-time utilization of BEJ task instances, which will incur high overhead and will be difficult to apply.

\subsubsection{Interference Scores of Servers} 
Similarly, we evaluate server interference predictions from three aspects: (1) the accuracy with an absolute error $\leq$ 2; (2) precision, recall, and F1-score for the BUSY category; and (3) severe prediction errors, especially the probability that a high interference level is predicted to be a low level. 
The four weighting schemes described in Section \ref{sec_score} are also evaluated. Figure \ref{ml} shows the prediction results of the 600 servers in our cluster.
The $Corr$ weighting scheme performs best as it directly reflects the correlation between the interference level and the CPU usage of BE instances. The performance of $Fair$ and $CPU$-Quota schemes is nearly identical, although the number of CPUs required by LCSs varies from 1 to 32 cores. $CV$ is not only related to BE pressure, but also to many other factors, such as the business logic of LCSs.

\section{Conclusion}
\label{sec-conclusion}
Interference mitigation in large-scale co-located clusters is challenging due to the high dynamics as well as efficiency and reliability requirements.
Based on the patterns extracted from our real production cluster, we propose PISM to achieve interference-aware scheduling for BEJs with a scoring and mitigating mechanism. PISM can be seamlessly integrated into any scheduler to mitigate interference.
Although we design PISM based on patterns extracted from our production cluster, it is a systematic and general approach that can be adopted in other clusters.
The simulation results based on the cluster trace show that PISM can effectively reduce the interference level by up to 41.5\%. In addition, PISM can reduce the probability of long-tail occurrence by 82.1\%. 

Since LCSs and BEJs co-run on the same server, when the interference of LCSs is reduced, BEJs benefit as well. We leave the detailed analysis of BEJ performance improvement as future research work to comprehensively investigate the interference problem in large-scale data centers.

\backmatter





\bmhead{Acknowledgements}

Acknowledgments are not compulsory. Where included they should be brief. Grant or contribution numbers may be acknowledged.

Please refer to Journal-level guidance for any specific requirements.

\section*{Declarations}

\begin{itemize}
\item Ethical Approval

We confirm that the study has not been published and is not under consideration for publication elsewhere. Further, this submission has been approved by all authors and the institution where the study was conducted (Alibaba Group and Shanghai Jiao Tong University), and we wish to be considered for publication in The Journal of Supercomputing.

\item Funding

Dingyu Yang is supported by the National Natural Science Foundation of China (No.61702320) and Jian Cao is supported by 
National Natural Science Foundation of China (No.62072301).

\item Availability of data and materials

The cluster data are available  and published at Github: 
(1) Alibaba Cluster (https://github.com/alibaba/clusterdata)
(2) Google Cluster (https://github.com/google/cluster-data)

\item Author contribution

Dingyu, Kangpeng, and Shiyou wrote the main manuscript text. Dingyu and  Kangpeng conducted the experiments and data validation. Shiyou and Jian Cao also designed the framework of PISM. Jian Cao and Guangtao Xue proofreaded our work. All authors reviewed the manuscript before submission.
\end{itemize}


\end{document}